\documentclass[manuscript,screen,nonacm]{acmart}
\AtBeginDocument{%
  }

\setcopyright{none}
\renewcommand\footnotetextcopyrightpermission[1]{}
\acmConference{}{}{}

\usepackage{longtable}
\usepackage{array}
\usepackage{enumitem}
\usepackage[most]{tcolorbox}
\usepackage{pdflscape}
\usepackage[most]{tcolorbox}

\usepackage{pgfplots}
\pgfplotsset{compat=1.18}
\usepackage{tikz}

\usepackage[most]{tcolorbox}   % 'most' loads breakable, skins, listings, etc.
 
\newtcolorbox{promptbox}[1][]{%
  enhanced,
  breakable,                    % lets the box split across columns/pages
  colback=black!4,
  colframe=black!65,
  boxrule=0.5pt,
  arc=1.5pt,
  left=4pt, right=4pt, top=3pt, bottom=3pt,
  fontupper=\footnotesize\sffamily\raggedright,  % small sans-serif, ragged-right avoids bad hyphenation in narrow columns
  parskip=3pt,
  before skip=6pt, after skip=6pt,
  title={System Prompt for Transcript Conversation Category Classification},
  fonttitle=\footnotesize\bfseries\sffamily,
  coltitle=white,
  colbacktitle=black!65,
  attach boxed title to top left={yshift=0pt},
  boxed title style={sharp corners, size=small},
  #1
}

\usepackage{booktabs}
\usepackage{multirow}
\usepackage[table]{xcolor}
\usepackage{pgf}
\usepackage{listings}
\newcommand{\heat}[1]{%
  \begingroup
  \pgfmathtruncatemacro{\shade}{%
    max(5,min(35,5+1.5*(#1-80)))%
  }%
  \edef\temp{\endgroup
    \noexpand\cellcolor{blue!\shade}%
  }%
  \temp
  #1%
}

\begin{document}
\title[Why LGBTQ+ People Choose (Not) to Use AI Chatbots for Personal Health and Relationship Advice]{``I had nobody to turn to'': Why LGBTQ+ People Choose (Not) to Use AI Chatbots for Personal Health and Relationship Advice}

\author{Jacob Hobbs}
\affiliation{
  \institution{Queer in AI \& Carnegie Mellon University}
  \country{USA}
  }
\authornote{First author, please send correspondence to jacobhobbs1100@gmail.com}
\email{jacobhobbs1100@gmail.com}

\author{Ashray Gupta}
\affiliation{
  \institution{Mahindra University}
  \country{India}
}
\authornote{Equal contribution, alphabetical order by first name}

\author{Dennis Ulmer}
\affiliation{
  \institution{Queer in AI \& University of Amsterdam}
  \country{Netherlands}
}
\authornotemark[2]

\author{Geetanjali Aich}
\affiliation{
  \institution{University of Massachusetts Amherst}
  \country{USA}
}
\authornotemark[2]

\author{Nolan Golden}
\affiliation{
  \institution{Queer in AI \& Carnegie Mellon University}
  \country{USA}
}
\authornotemark[2]

\author{Yanan Long}
\affiliation{
  \institution{Queer in AI \& StickFlux Labs}
  \country{UK}
}
\authornotemark[2]

\author{Ian Rios-Sialer}
\affiliation{
  \institution{Queer in AI}
  \country{USA}
}
\authornote{Equal contribution, alphabetical order by first name}

\author{Jennifer Mickel}
\affiliation{
  \institution{Queer in AI \& UC Berkeley}
  \country{USA}
}
\authornotemark[3]

\author{Morgan Klaus Scheuerman}
\affiliation{
  \institution{Queer in AI}
  \country{Spain}
}
\authornotemark[3]

\author{Rohit Avinash Mujumdar}
\affiliation{
  \institution{Queer in AI}
  \country{USA}
}
\authornotemark[3]

\author{Virginia K. Felkner}
\affiliation{
  \institution{University of Southern California Information Sciences Institute \& Queer in AI}
  \country{USA}
}
\authornotemark[3]

\author{Jordan Taylor}
\affiliation{
  \institution{Queer in AI \& University of Michigan}
  \country{USA}
}
\authornote{Advisor}

\author{William Agnew}
\affiliation{
  \institution{Queer in AI \& Carnegie Mellon University}
  \country{USA}
}
\authornotemark[4]

\renewcommand{\shortauthors}{Hobbs et al.}
\authorsaddresses{}
\begin{abstract}

    While there is an abundance of research into large language model (LLM) biases against LGBTQ+ communities, far less research has examined how and why LGBTQ+ people choose to make use of AI chatbots.
    In this work, we examine how LGBTQ+ people choose (not) to use AI chatbots across three domains in which LGBTQ+ users have unique needs: relationship support, mental health support, and gender or sexual health support. 
    We surveyed 73 people (60 LGBTQ+; 13 non-LGBTQ+) on how they interact with AI chatbots and find that participants' access to external resources, privacy concerns, and perceptions of chatbot capabilities shape whether and how they choose to make use of AI chatbots. 
    For instance, some LGBTQ+ participants felt that chatbots might actually be \textit{less biased} against LGBTQ+ people than the people they know offline. We also find that LGBTQ+ participants report using AI chatbots less frequently than non-LGBTQ+ participants across 15 specific personal health and relationship advice scenarios, citing concerns about AI sycophancy and adverse cognitive side-effects.
    Recognizing AI chatbots as vital (albeit imperfect) resources for LGBTQ+ people, we recommend researchers consider systemic issues that drive users to AI chatbots and design technologies (AI or otherwise) that prioritize privacy, anonymity, and real human connection. 

\end{abstract}

\maketitle
\fancypagestyle{preprint}[standardpagestyle]{%
  \fancyfoot{}%
  \fancyfoot[L]{\footnotesize Preprint}%
}
\fancypagestyle{preprintfirst}[firstpagestyle]{%
  \fancyfoot{}%
  \fancyfoot[L]{\footnotesize Preprint}%
}
\thispagestyle{preprintfirst}
\pagestyle{preprint}

\section{Introduction}
A 2026 report by GLAAD, a leading LGBTQ+ advocacy organization, characterizes artificial intelligence (AI) as having the ``potential to deliver meaningful benefits for LGBTQ people'', but emphasizes that ``LGBTQ people, like other historically marginalized communities, face unique impacts and potential real-world harms from the ways AI models and products are designed and used'' \cite{glaad2026lgbtqimpacts}. This report is especially important at a time when AI products are increasingly integrated into everyday life. Many users rely on AI systems for sensitive tasks such as mental health support, inter-personal advice, and diagnostic purposes \cite{lissak_colorful_2024, aiobservatory2026}. Understanding how users interact with these systems is critical to crafting effective evaluations and mitigation strategies to improve existing AI systems, particularly for sensitive tasks, high-risk domains, and historically marginalized communities. Although prior research has shown that large language models (LLMs) can produce heteronormative or stereotypical outputs in response to LGBTQ+ community language \cite{felkner_winoqueer_2023, tint_guardrails_2025, hirsch_implicit_2026}, there is still a need for better understanding of how LGBTQ+ people are actually using AI products. This understanding of how specific communities choose to interact with technology is critical to reduce harm from existing systems and to inform community-centered design of new ones.

\begin{figure*}
    \centering
    \includegraphics[width=0.985\textwidth]{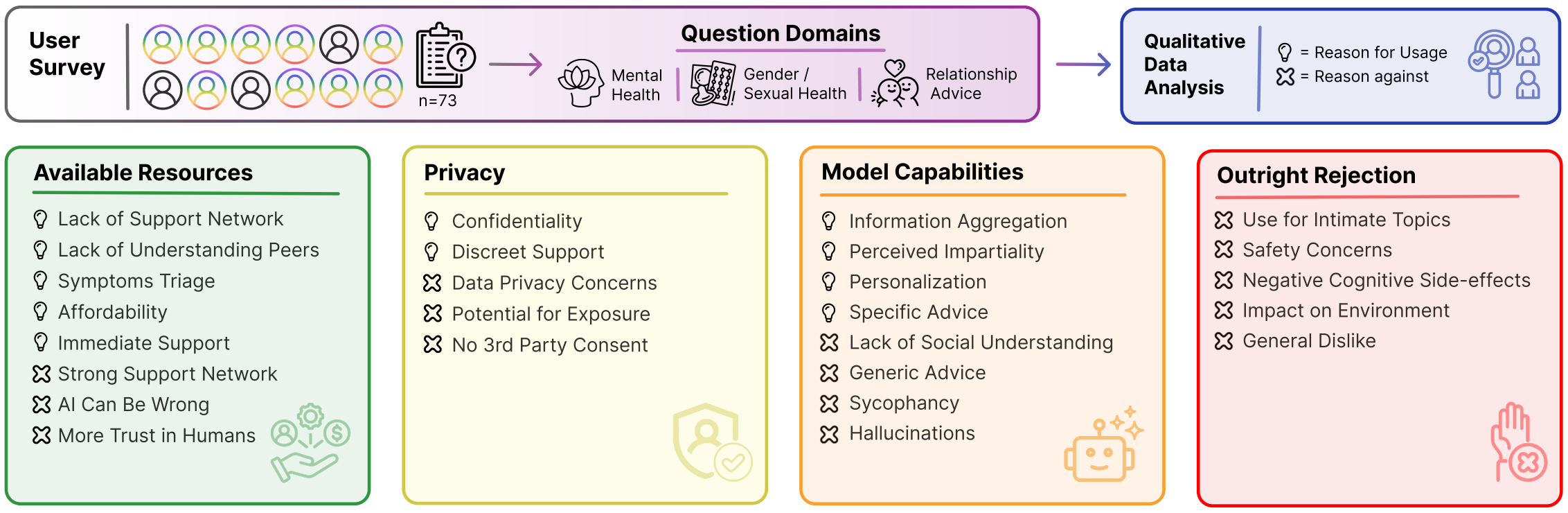}
    \caption{Our user survey focuses on chatbot use for questions around mental health, gender and sexual health, and relationship advice. The study consisted of $n=73$ participants, with $82 \%$ identifying as LGBTQ+. Through qualitative data analysis, we uncover distinct reasons for chatbot use in the LGBTQ+ community, primarily stemming from resource access, privacy concerns, model capabilities, and reasons for outright rejection.}\label{fig:overview}
\end{figure*}

In this work, we investigate the specific ways LGBTQ+ users interact with AI chatbots when asking for advice on mental health, gender and sexual health, and relationship advice. We selected these domains because LGBTQ+ people have unique health \cite{liang2026scoping} and relationship \cite{kinnee2022sonic} needs that differ from non-LGBTQ+ people. Much like LGBTQ+ communities have long used online communities for informational and emotional support \cite{fox2016queer, gray2009negotiating, dym2019coming, haimson2021tumblr}, nascent research suggests that LGBTQ+ people are also increasingly turning to AI chatbots \cite{liem_enhancing_2026, ma_evaluating_2024}. At the same time, LGBTQ+ communities have not universally embraced AI \cite{taylor_i_2026, haimson_ai_2025}. For instance, \citet{haimson_ai_2025} find that non-binary and transgender people have more negative attitudes toward AI, broadly defined, than cisgender people. We expand on this work by studying how and why LGBTQ+ people choose (not) to use AI chatbots for support in the health and relationship domains. We do so through an online survey that aimed to answer the following research questions:

\begin{enumerate}
    \item[\textbf{RQ1}] Why, and under what conditions, would LGBTQ+ users turn to an AI chatbot over a conventional resource?
    \item[\textbf{RQ2}] What concerns do LGBTQ+ chatbot users have with AI chatbots?
    \item[\textbf{RQ3}] What factors cause LGBTQ+ people to avoid using AI chatbots?
\end{enumerate}

To address these questions, we conducted an online survey of 73 people (60 LGBTQ+; 13 non-LGBTQ+) on how they interact with AI chatbots across mental health, gender and sexual health, and relationship advice domains. 
Through a mixed-method analysis of survey responses, we found that three key factors mediate how LGBTQ+ respondents use, or do not use, AI chatbots: \textbf{available resources} (e.g., does the participant have a supportive social network or financial resources to afford professional help?), \textbf{perceptions of privacy} (e.g., does the participant have concerns about the handling of their sensitive data, or do they seek a private confidant in AI?), and \textbf{perceptions of model capabilities} (e.g., does the participant think AI can actually answer their question, provide an unbiased answer, and avoid sycophancy?). 
Additionally, we found that LGBTQ+ participants report using AI chatbots 
less frequently than non-LGBTQ+ participants across 15 specific scenarios. Finally, consistent with \citet{haimson_ai_2025}, we found that transgender respondents felt much less positive about AI chatbots in general compared to other respondents. 
An overview of the findings of our qualitative analysis is given in Figure \ref{fig:overview}. In light of our findings, we encourage AI designers and policymakers to:

\begin{enumerate}
    \item Design AI chatbots with LGBTQ+ privacy in mind
    \item Design AI chatbots to build (rather than replace) LGBTQ+ social information seeking
    \item Recognize AI chatbots as vital (albeit imperfect) resources for LGBTQ+ people
\end{enumerate}

\section{Related Work}

In this section, we first ground our work in literature exploring LGBTQ+ community relationships with AI systems. Then, we overview the politics of online information seeking.

\subsection{Queer AI Research}

There is a growing volume of research on how LGBTQ+ people (don't) use AI systems and how these systems tend to reflect homophobic, transphobic, or heteronormative trends in training corpora. Researchers have studied how and if LGBTQ+ communities choose to engage with AI systems in domains such as health \cite{ma_evaluating_2024, liem_enhancing_2026, buslon_exploring_2025}, art \cite{taylor_i_2026, taylor_-straightening_2025}, and voice training \cite{rincon_speaking_2021}. % For LGBTQ+ experiences with AI, there is some prior work with LGBTQ+ focus groups on AI-assisted health support \cite{ma_evaluating_2024, liem_enhancing_2026, buslon_exploring_2025}, art generation \cite{taylor_i_2026, taylor_-straightening_2025}, and audio domains \cite{rincon_speaking_2021}. 
There is also evidence to suggest that LGBTQ+ people might hold more negative attitudes toward AI than more dominant groups:  \citet{haimson_ai_2025} and \citet{ungless_amplifying_2025} found that non-binary and transgender individuals generally have more negative views on AI.  
For evaluations of language technology, prior work has shown that mentioning LGBTQ+ language decreases model performance \cite{testoni_calibrated_2026}, disproportionally triggers AI content moderation systems \cite{dias_oliva_fighting_2021, monea_digital_2023}, is filtered out from training corpora \cite{dodge_documenting_2021, taylor2026algorithmic}, and produces overall biased output \cite{felkner_winoqueer_2023, tint_guardrails_2025, hirsch_implicit_2026}. Audits have also found that AI systems produce constrained or stereotypical depictions of LGBTQ+ identity across text \cite{gillespie_generative_2024, ostrow_llms_2025}, image \cite{ungless_stereotypes_2023}, and audio \cite{weber_sounds_2026} tasks. \citet{scheuerman_transphobia_2025} find that LLMs can seem trans affirming, while still perpetuating more subtle forms of transphobia. \citet{weber_queer_2026} and \citet{hobbs_theories_2026} have also conducted audits of how LGBTQ+ identities are operationalized in natural language processing (NLP) research, finding that NLP bias researchers often do not employ participatory methods with the LGBTQ+ community or ground their work in relevant social theory \cite{weber_queer_2026, hobbs_theories_2026}. We expand on this existing research that documents the complex relationship between AI chatbots and the LGBTQ+ community by taking a bottom-up approach to understanding LGBTQ+ attitudes towards AI systems in particularly sensitive health and interpersonal advice domains.

\subsection{Chatbots and Information Seeking}

Foundational work on how users interact with human support networks versus algorithmic search engines can offer insight into how AI positions itself as a unique source of knowledge. \citet{pirolli_elementary_2009} argues that users searching for information value the diversity of connections that social information retrieval systems offer over systems that are less socially connected. Similarly, \citet{chi_information_2009} explores how systems that feature social connection increase fact-finding effectiveness and enable finding useful related information along the way. Compared to search engines, online forums offer users greater social connectedness, more personalized answers on opinion-based questions, and novel perspectives that cannot be found elsewhere online \cite{morris_comparison_2010}. Online forums that satisfy users' need for emotional support also garner higher retention rates than sites that simply answer fact-based information \cite{wang_stay_2012}. However, algorithmic informational retrieval systems such as search engines enable speedy answers, iterative control over asked questions, and a comforting sense of perceived objectivity from users \cite{morris_comparison_2010}.

Traditional online information retrieval systems also offer unique opportunities for users to co-construct their sense of self through community-centered social forums. This affordance is even more impactful for users of marginalized communities who lack effective support networks in their everyday lives. \citet{fricker_epistemic_2007} coined the term epistemic injustice to describe how members of these groups often lack access to shared knowledge or common terms to describe their experiences. Along these lines, \citet{taylor_mitigating_2024} explored how users on the Reddit page r/bisexual mitigate epistemic injustice by mutually constructing knowledge of who they are and how they can collectively navigate a shared disconnect against normative categorization. Similar research has explored how transgender communities share and form knowledge through online forums \cite{deguia_bridging_2026}, and how parents of LGBTQ+ youth re-construct  their own identity as supportive parents through relational connections to LGBTQ+ realities \cite{kwon_journeys_2026}. 

In recent years, AI chatbots have grown in popularity as a tool for information seeking \cite{chatbot_info_seeking, aiobservatory2026}. AI chatbots are often anthropomorphically designed to resemble human interlocutors \cite{devrio2025taxonomy}, and researchers increasingly use chatbots to simulate human participants \cite{kapania2025simulacrum, karetnikov_large_2026}. This quasi-humanness places contemporary AI chatbots somewhere in-between information seeking through search engines versus through embodied people in online communities. Research in this domain has shown that while AI-assisted chatbots can offer tailored advice at the speed of a search engine, they can create echo chambers that prioritize user retention over accurate output \cite{sharma_generative_2024}. Simultaneously, personalization features can produce biased output when models pick up on explicit and implicit user-identity cues \cite{kantharuban_stereotype_2025}, and LGBTQ+ users have expressed how LLMs can fail to grasp the nuances of their identity \cite{ma_evaluating_2024}. Even so, users are increasingly employing AI chatbots as information retrieval tools for subjective personal questions \cite{cheng_llms_2026} and in the sensitive domains of relationship advice \cite{tseng_chat_2026}, mental health \cite{moore_characterizing_2026, moore_expressing_2025, chen_empirical_2026}, and gender or sexual health topics \cite{liem_enhancing_2026, buslon_exploring_2025}. 

More than three decades before the widespread use of AI chatbots as information retrieval tools, feminist science scholar \citet{haraway_situated_1988} argued that there are no neutral sources of knowledge. In a critique of science's perceived objectivity, she argued that all knowledge is ``situated,'' meaning it is produced by individuals in particular cultural and historical contexts. Those with multiple minority identities, for example, produce knowledge that is informed by the effects of being multiply marginalized \cite{hill_collins_black_2002}. Rooted in these feminist theories that question the possibility of a ubiquitous objectivity, \citet{dignazio_data_2020} argue that data always comes from somewhere. The perceived neutrality of data can perpetuate what Haraway calls the ``god trick,'' or the illusion of ``seeing everything from nowhere,'' in which a particular, embodied perspective presents itself as disembodied and universal \cite{haraway_situated_1988}. This illusion is especially powerful for data-driven tools that present themselves as objective, hiding the real people and subjective contexts that formed them \cite{dignazio_data_2020}. Our work expands on these critiques by examining how users of AI systems recognize and navigate AI's anthropomorphic output that give strikingly human-like qualities to ostensibly ``objective'' machine output.

% ** end with how AI is trained on this potentially biased data so this is important...

\section{Methodology}

\subsection{Online Survey}

Our survey focused on mental health, gender and sexual health, and relationship advice -- three domains where LGBTQ+ people have unique needs \cite{liang2026scoping, kinnee2022sonic}. We chose these areas based on an author’s expertise working with LGBTQ+ people in community mental health centers, prior work showing LGBTQ+ people are increasingly using chatbots for these topics \cite{moore_characterizing_2026, tseng_chat_2026, liem_enhancing_2026}, and our lived experiences as queer researchers. To ground our survey questions (detailed in Appendix~\ref{appdx:example_survey_questions} and \ref{situational_question_topics}), we adapted items from validated minority stress measures, such as the Gender Minority Stress and Resilience (GMSR) Measure \cite{testa2015gender}, the Daily Heterosexist Experiences Questionnaire (DHEQ) \cite{balsam_daily_2013}, and the LGBT People of Color Microaggressions Scale \cite{balsam2011measuring}, alongside prior research on LGBTQ+ chatbot usage from \citet{ma_evaluating_2024}.

After selecting these domains, we iteratively developed a survey with a mix of Likert scale and free-response questions. Our survey aimed to collect the following key types of information to answer why LGBTQ+ participants turn to AI chatbots (\textbf{RQ1}) and what specific factors dissuade LGBTQ+ chatbot use (\textbf{RQ2} and \textbf{RQ3}):

\begin{enumerate}
    \item Basic demographic information and detailed gender/sexual identity information (Table~\ref{tab:demographics}).
    \item How often respondents use AI chatbots in general and for advice in our three focus areas (Appendix~\ref{usage_frequency}).
    \item How comfortable respondents feel mentioning their identity to an AI chatbot (Appendix~\ref{identity_disclosure}).
    \item How respondents feel about using AI chatbots in various contexts (Appendix~\ref{attitute}).
    \item How respondents have used, would use, or would not use AI chatbots in 15 specific scenarios relating to our three focus areas (Appendix~\ref{example_question_structure} \& \ref{situational_question_topics}).
\end{enumerate}

\subsubsection{Survey distribution}
\label{method:distribution}

In order to reach a varied sample of respondents, we utilized the survey distribution platform Prolific and applied various screening criteria. Respondents had to be fluent in English and at least 18 years old. Additionally, we set quotas to achieve a 75\% total response rate from LGBTQ+ respondents and at least a 20\% total response rate from transgender respondents. We wanted to include some ($\sim$25\%) cisgender/heterosexual participants to compare responses and potentially notice differences in attitudes/uses of AI chatbots compared to LGBTQ+ participants. In a similar vein, we over-sampled for people who identify as transgender to guarantee diverse gender identity representation. Our first phase of data collection yielded a few United States respondents, but their responses were notably higher in quality than non-US responses. We also did not meet our goal of 20\% transgender respondents after this first phase. 
Therefore, we decided to further restrict responses to only current US residents and to refine our filtering to more effectively reach transgender participants.
More details about survey distribution are described in Appendix~\ref{appx_distrubution_method}. The survey was designed to take the average responder about 30 minutes to complete; the average response time was 34.6 minutes (\textit{SD} = 17.7, range = 6.3–100.0). Each participant was compensated \$10 USD for completing the survey.

\subsubsection{Participants}
\label{method:participants}

Our survey ran in August and September 2026. 118 participants agreed to take part in the study; out of these, 100 participants completed the survey, and 18 either timed out or withdrew. After reviewing these initial 100 survey responses, two researchers collectively discarded 21 responses due to signs of bot activity or poor quality (e.g., overly quick responses). We also decided to exclude responses that did not primarily respond in English ($n=6$)\footnote{These participants were removed after assigning participants internal IDs, meaning some participants cited throughout the paper have IDs > 73}. After this, we were left with 73 survey responses (see Table~\ref{tab:demographics} for demographics). 

\begin{table}[t]
  \caption{Participant demographics ($n=73$). Gender, sexuality, and
  race/ethnicity were multi-select, so counts exceed $n$ and percentages
  sum to more than 100\%. All percentages are of $n=73$ and are rounded
  to the nearest whole number. Write-in responses under \textit{Prefer to
  self-describe} (gender) were ``Questioning'' and ``Gender queer.''}
  \Description{A three-part table of participant demographics. The left
  part reports LGBTQ+ identity, transgender identity, disability, and age.
  The middle part reports sexuality and gender. The right part reports
  race/ethnicity and country of residence.}
  \label{tab:demographics}
  \small
  \newcommand{\fitcol}[1]{\resizebox{\ifdim\width>\linewidth\linewidth\else\width\fi}{!}{#1}}
  \begin{minipage}[t]{0.30\textwidth}
    \centering
    \fitcol{%
    \begin{tabular}[t]{@{}lrr@{}}
      \toprule
      \textbf{Characteristic} & \textbf{n} & \textbf{\%} \\
      \midrule
      \multicolumn{3}{@{}l}{\textit{LGBTQ+}} \\
      \quad Yes & 60 & 82 \\
      \quad No  & 13 & 18 \\
      \addlinespace
      \multicolumn{3}{@{}l}{\textit{Transgender}} \\
      \quad No  & 52 & 71 \\
      \quad Yes & 21 & 29 \\
      \addlinespace
      \multicolumn{3}{@{}l}{\textit{Disability}} \\
      \quad No & 52 & 71 \\
      \quad Yes & 18 & 25 \\
      \quad Prefer not to answer & 3 & 4 \\
      \addlinespace
      \multicolumn{3}{@{}l}{\textit{Age}} \\
      \quad 18--24 & 17 & 23 \\
      \quad 25--34 & 35 & 48 \\
      \quad 35--44 & 17 & 23 \\
      \quad 45--54 & 3 & 4 \\
      \quad 55--64 & 1 & 1 \\
      \bottomrule
    \end{tabular}%
    }
  \end{minipage}
  \hfill
  \begin{minipage}[t]{0.30\textwidth}
    \centering
    \fitcol{%
    \begin{tabular}[t]{@{}lrr@{}}
      \toprule
      \textbf{Characteristic} & \textbf{n} & \textbf{\%} \\
      \midrule
      \multicolumn{3}{@{}l}{\textit{Sexuality} (multi-select)} \\
      \quad Bisexual & 28 & 38 \\
      \quad Straight or heterosexual & 14 & 19 \\
      \quad Gay & 11 & 15 \\
      \quad Pansexual & 10 & 14 \\
      \quad Lesbian & 9 & 12 \\
      \quad Queer & 8 & 11 \\
      \quad Asexual & 7 & 10 \\
      \quad Questioning & 3 & 4 \\
      \quad Demisexual & 2 & 3 \\
      \addlinespace
      \multicolumn{3}{@{}l}{\textit{Gender} (multi-select)} \\
      \quad Man & 33 & 45 \\
      \quad Woman & 32 & 44 \\
      \quad Non-binary & 15 & 21 \\
      \quad Prefer to self-describe & 2 & 3 \\
      \bottomrule
    \end{tabular}%
    }
  \end{minipage}
  \hfill
  \begin{minipage}[t]{0.34\textwidth}
    \centering
    \fitcol{%
    \begin{tabular}[t]{@{}lrr@{}}
      \toprule
      \textbf{Characteristic} & \textbf{n} & \textbf{\%} \\
      \midrule
      \multicolumn{3}{@{}l}{\textit{Race/ethnicity} (multi-select)} \\
      \quad White/Caucasian & 47 & 64 \\
      \quad Black/African/African American & 19 & 26 \\
      \quad Latino & 5 & 7 \\
      \quad Hispanic & 4 & 5 \\
      \quad Prefer to self-describe & 4 & 5 \\
      \quad Native American & 3 & 4 \\
      \quad Middle Eastern & 2 & 3 \\
      \quad North Asian & 1 & 1 \\
      \quad Southeast Asian & 1 & 1 \\
      \quad South Asian & 1 & 1 \\
      \quad Pacific Islander & 1 & 1 \\
      \addlinespace
      \multicolumn{3}{@{}l}{\textit{Country of residence} (17 total)} \\
      \quad United States & 28 & 38 \\
      \quad South Africa & 12 & 16 \\
      \quad United Kingdom & 8 & 11 \\
      \quad Brazil & 4 & 5 \\
      \quad Portugal & 4 & 5 \\
      \quad Other (12 countries) & 17 & 23 \\
      \bottomrule
    \end{tabular}%
    }
  \end{minipage}
\end{table}

\subsection{Data analysis}

Our survey collected various types of data, ranging from Likert-scale questions on perceptions of AI chatbots to free-text questions that detail specific AI chatbot (non) use.
Prior to analyzing this data, our research team removed all personally identifiable information (PII) through automatic replacement via Presidio \cite{MsPresidio} and manual verification. We then qualitatively analyzed each participant’s open-ended responses to characterize their specific motivations for using AI chatbots. To support these qualitative findings, we ran statistical tests on the closed-ended survey questions.

\subsubsection{Qualitative data analysis}

Each of our quantitative survey questions was accompanied by an optional, open-ended text field in which participants could explain their feelings toward and experiences using AI chatbots. The first and last authors first familiarized themselves with the data through the aforementioned data quality checks. Then, nine members of our research team independently open-coded the free response fields from the survey \cite{corbin_basics_2015}, resulting in 1,852 codes. Our qualitative data analysis lasted approximately two months, during which our research team met bi-weekly to discuss shared patterns. Following our initial open coding, the first author exported each code into a digital whiteboarding tool, which we used to collectively conduct axial coding during group meetings. At this point, we began to connect AI chatbot usage to broader, structural factors in participants' lives that mediate chatbot usage, such as access to LGBTQ+ affirming social connections. We discuss these patterns in greater detail in Section~\ref{sec:findings}.

\subsubsection{Structured data analysis}
We used Bayesian hierarchical regression to analyze the responses from participants on willingness and frequency of chatbot usage across the 15 scenarios, grouped by our three focus areas (mental health, gender and sexual health, and relationship advice). We structured the comparisons in three ways: LGBTQ+ vs non-LGBTQ+, transgender vs non-transgender, transgender vs non-transgender LGBTQ+. 
We modeled actual use and willingness as binary variables parameterized by a linear model with user- and question-specific slopes and offsets.
We treat usage frequency within a given time frame as an ordinal variable given the Likert-style responses (see Appendix~\ref{usage_frequency}), and model the probability of the number of uses falling below a specific threshold.
We fitted all outcomes in a joint model, with participant-specific offsets capturing the fact that some respondents use chatbots more than others overall.
We report results as differences in percentage points (pp) in probability (relative difference or RD), where a value above $0$ suggests a higher use / willingness / usage frequency in LGBTQ+ and trans-identified respondents than non-queer respondents.
We also report the probability that the difference is less than $0$, denoted by $\text{Pr}(< 0)$.\footnote{
    We report these pieces of information rather than binary significance tests which would not show effect sizes, based on common critiques of statistical significance testing, e.g.\@ \citep{johnson1999insignificance, kruschke2013bayesian, wasserstein2019moving, azer2020not}.
    
}
Results from our statistical analysis are detailed in the beginning of Section~\ref{sec:findings}, with additional model specifications and result tables reported in Appendix~\ref{appdx:stats}.

\section{Findings}
\label{sec:findings}
\begin{figure*}[t]
    \centering
    \includegraphics[width=0.96\textwidth]{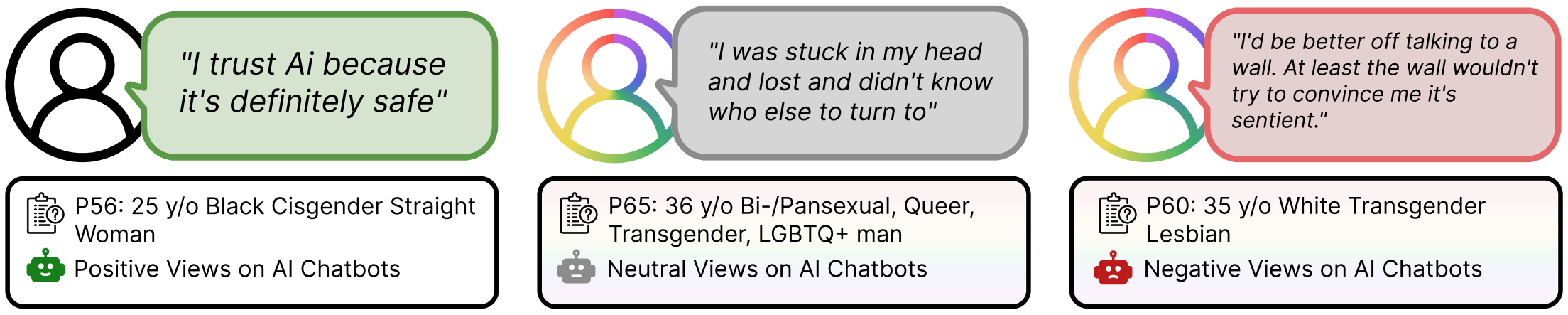}
    \caption{Examples of participants and their stated attitudes towards AI chatbots. In our responses, we find non-LGBTQ+ participants generally have more positive attitudes of AI (P56) compared to LGBTQ+ participants (P60). Some LGBTQ+ participants expressed mixed views on AI chatbots when they perceived a lack of alternative outlets to address their needs (P65).}
    \label{fig:three-participant-examples}
\end{figure*}

LGBTQ+ participants in our study used AI chatbots less than non-LGBTQ+ people. Specifically, we found that $ 19.8\% $ of LGBTQ+ participants and $17.8\%$ of trans respondents reported having actually used an AI chatbot in at least one of the 15 scenarios, compared to $ 29.4\%$ of non-LGBTQ+, non-trans respondents.
Averaging across all 15 scenarios, LGBTQ+ participants were $8.5\%$ less likely ($[-18.6, 0.5] $, $ \mathrm{Pr}(<0) = 96.7\%)$ than non-LGBTQ+ participants to have used a chatbot for a given scenario, with a smaller effect for trans compared to non-trans respondents ($5\%$ less likely, $ [-12.4, 2.5] $, $\mathrm{Pr}(<0) =  90.5 \% $).
The difference between LGBTQ+ and non-LGBTQ+ was clearest for the relationship domain, where LGBTQ+ participants were on average  $10.0 \% $ ($ [-21.7, -0.4] $, $\mathrm{Pr}(<0) = 98.0 \% $) less likely to have used a chatbot than non-LGBTQ+ participants. The $95\%$ credible interval for this comparison does not include zero, and the probability that this difference is less than zero is $98.0\%$. The difference between LGBTQ+ and non-LGBTQ+ respondents is also quite clear in the mental health domain, where LGBTQ+ users are $10.0\%$ ($ [-22.8, 2.1] $, $\mathrm{Pr}(<0) = 94.5 \% $) less likely to have used chatbots, although the credible interval overlaps zero, so this effect size is not quite as strong. In the sexual health domain, LGBTQ+ users are only $5.4\%$ ($ [-14.3, 2.0] $, $\mathrm{Pr}(<0) = 90.8 \% $) less likely to report past AI use, and the probability that the true difference is less than zero is only $90.8 \%$. Overall, we see statistical evidence that there is a small but notable gap in reported chatbot use between LGBTQ+ and non-LGBTQ+ respondents. At the same time, we cannot establish a direct causal link between these statistical results and the respondents' answers. 

In order to understand \textit{why} (or why not) LGBTQ+ people choose use AI chatbots, this section draws on our qualitative analysis of free-text survey responses (e.g., Figure~\ref{fig:three-participant-examples}). We found three major reasons why our LGBTQ+ participants chose (not) to use AI chatbots, which we detail below: (1) access to external resources, (2) privacy perceptions, and (3) perceptions of model capabilities. In contrast to those carefully choosing how or when to use chatbots, our last findings section describes why a subset of participants chose to reject chatbots in all circumstances.

\subsection{How Participants' Available Resources Shape Chatbot (Non) Use}
\label{findings:avail-resource}

We found that participants' access to social, informational, and financial resources majorly contributed to whether they sought help from an AI chatbot or a conventional resource (e.g., a friend, medical professional, or online forum).  At the same time, even though some LGBTQ+ participants expressed that they had ample access to social and informational outlets, AI chatbots acted as a viable alternative when these resources didn't meet their specific needs.

\subsubsection{Social Resources}
\label{sec:social-resources}

Across mental health and relationship advice questions, users cited preference for human support, connection, and lived experience as a significant reason why they would \textit{not} use AI chatbots for subjective topics dealing with the ``human condition'' (P33) or ``human behavior'' (P31). These participants cited friends (P33, P50, P58, P61), family (P27, P32, P33), significant others (P28, P31), and their general community (P33, P61) as trusted sources of support when dealing with these topics. For example, one bisexual respondent with neutral feelings about AI in general but strong negative views on using AI for mental health expressed they would not consider seeking support for mental health issues because \textit{``[AI is] not a real person, I have friends and family that can help me better with that''} (P33). 

Although some participants did not want to use AI because they preferred human support, LGBTQ+ participants who lacked a strong support network did not have these human alternatives. Among these participants, a common response to why they \textit{would} or \textit{did} reach out to an AI chatbot across scenarios is because they felt there was simply nobody else they could turn to in a time of need (P1, P57, P69). Reliance on AI chatbots for support seemed to be compounded when LGBTQ+ participants felt questions would not be received or understood well by typical support structures:

\begin{quote}
    \textit{``My friends do not understand or identify as transgender. They are also thin and lean.''} (P59)
    
    \textit{``There is a bit of stigma in my community about mental health issues''} (P30)

    \textit{``I can go to AI for coping mechanism as someone dealing with society rejecting my sexuality''} (P2)
\end{quote}
In the mental health domain, users depend on AI chatbots for immediate support in times of desperation or crisis (P22, P65), with one participant mentioning they were ``\textit{suicidal}'' and AI ``\textit{gave [them] many ways to seek for help}'' (P2). 
 Some participants felt that, in sensitive areas like mental health, AI was a suitable, empathetic alternative in the absence of genuine human connection (P15, P42, P59).
Participants viewed AI as providing critical resources when they lacked reliable, understanding social networks in times of desperate need.

\subsubsection{Informational Resources}
\label{findings:informational-resources}
Participants who were hesitant to use AI chatbots for medical and fact-based queries often cited greater trust in medical professionals or traditional online resources (e.g., Google, online forums) when accuracy was important (P1, P42, P62, P66, P67). The consequences of potentially wrong recommendations about gender and sexual health medications seemed to dissuade one non-binary, demisexual, and pansexual participant from using AI: ``\textit{I can research this on my own using trustworthy medical sources, since this is a big deal}'' (P1).
However, the immediate and tailored nature of AI responses was alluring for participants who used chatbots for triaging medical questions. Participants valued AI's ability to assess if their specific symptoms were serious enough to seek professional help (P7, P28, P58, P63) and how AI made them feel more prepared to discuss treatment with professionals (P14, P21, P52, P63). 

Participants generally preferred conventional, non-AI resources for personal health questions, but when LGBTQ+ users lacked reliable access to understanding professionals or communities of care, AI served as a critical source of information. One participant expressed that AI offered an alternative for him ``\textit{[m]erely because [he doesn't] have access to real medical professional help}'' (P37). Some LGBTQ+ users turned to AI because their immediate communities were ignorant about, or even antagonistic toward, issues like gender affirming care or mental health medication (P30, P52). P77, a transgender man who had questions about the timing of bottom growth while on testosterone, expressed that ``\textit{[o]nline sources tend to have different answers and [he] want[ed] an easy overview}.'' In this way, AI's ability to aggregate, synthesize, and recall specific knowledge serves an important role for LGBTQ+ participants who feel that online or in person sources are inadequate for their needs. This is discussed further in Section~\ref{sec:the-politics-of-ai}.

\subsubsection{Financial Resources}

A strong reason why participants used AI chatbots was affordability (P4, P20, P28, P42, P53, P58, P59, P69, P73). For many participants, AI serves as a last-ditch alternative when typical resources are prohibitively expensive. These participants indicated they would prefer to see a professional if they had the financial means to do so. P58 explained: \textit{``[AI is] free, full stop, I'd rather an actual therapist in all cases}'' (P58). Likewise, P42 used AI chatbots for mental health support because they ``\textit{have no money to pay for professional help}.'' Participants who mentioned lack of access to external social and informational resources often cited affordability as an additional reason for using AI.

\subsection{How Participants' Privacy Concerns Shape Chatbot (Non) Use}

We found LGBTQ+ respondents held a wide range of opinions on whether to discuss sensitive, personal information with AI chatbots. Participants usually framed these opinions in terms of privacy: some were concerned about the collection of their personal data, while others felt like talking to a chatbot was \textit{more private} than talking to a person. 

\subsubsection{AI chatbots provide a safe space for users discussing sensitive or embarrassing topics}
\label{findings:safe-space}

Participants felt comfortable using an AI chatbot for support across our three domains partially because the chatbots felt more private than talking to a person. These participants praised the perceived sense of confidentiality (P47), privacy (P69), and anonymity (P74) that chatbots offered. Participants who used chatbots as a private confidant felt that AI protected them \textit{from} criticism, fearing that humans would judge their situation (P20, P57, P68),  ``berate'' or ``mock'' their opinions (P6, P59), or label their viewpoint ``politically incorrect'' (P25). Some LGBTQ+ users specifically appreciated that AI can offer human-like support and a private ear when they are not yet public about their identity  (P30, P68). One of these participants (P68) explained that they were ``\textit{still [in] the closet at the time}'' as to why they asked AI, instead of a friend or another resource, about how their sexual orientation may change after transitioning. Another used Google's Gemini to explore the possibility of ``\textit{being happy with a family}'' as a gay man struggling with dominant heteronormative narratives surrounding marriage (P53). These users sought a safe space from the outside world and from human judgment, and AI offered an ostensibly protective outlet.

\subsubsection{Concerns about individual data privacy}
\label{findings:data-privacy}

Some LGBTQ+ participants were particularly concerned about how their personal data would be handled by AI companies. Specifically,  participants were hesitant to disclose their LGBTQ+ identity (P3, P61, P69, P72) or healthcare information (P72) to AI chatbots due to general privacy concerns (P61, P74), distrust about the security of their data (P5, P78), and the fear that their data could end up in the wrong hands (P69). One participant who otherwise felt positively about using AI chatbots explained he would not discuss his gender/sexual orientation:
\begin{quote}
    \textit{``I'd be scared [the AI chatbot] was somehow hooked up to the government trying to track transgender people. Conspiracy stuff I know but with the recent attitudes towards trans people i try not to leave a record''} (P69)
\end{quote}
These concerns reveal that both participants' trust and distrust of AI's privacy are driven by similar desires to avoid exposure. This creates a tension: LGBTQ+ users who value AI as a refuge from human scrutiny, such as those not yet out about their identity, are also the users that stand to risk the most from a data break or involuntary disclosure. For P69, leaving a ``record'' of his identity could mean involuntary outing.

\subsubsection{Privacy dynamics change when third parties are involved}

We found that participants were more concerned about privacy issues in the relationship advice domain. These concerns took two forms: acknowledged lack of consent from a third party and perceived consequences for participants if a third party found out about AI use. For the former, one participant expressed discomfort with using AI tools to seek communication or intimacy advice because they felt it would be ``\textit{invading the other person's consent}'' (P6). For the latter, three participants noted that if a partner used AI for relationship advice, they would have fundamental doubts about the integrity of the relationship itself (P10, P59, P61). One participant noted ``\textit{AI has no place in relationships. If I found out my partner was using AI to navigate our relationship, I would immediately break up with them}'' (P61). While some users did express they have used AI to resolve relationship and interpersonal issues, there was a shift in concern when participants considered the social consequences in potentially breaching \textit{another person's} privacy through an AI chatbot.

\subsection{How Participants' Perceptions of Model Capabilities Shape Chatbot (Non) Use}

Participants' understanding of AI chatbot features and behaviors had a strong effect on their willingness to use a chatbot for a sensitive query. We found that four perceived aspects of model capabilities were especially influential: extent of chatbot knowledge, (im)partiality, personalization, and sycophancy. These factors describe how users understood and valued AI's supposed expertise, potential biases, capacity to give specific advice, and general agreeableness.

\subsubsection{Extent of knowledge}
\label{findings:extent-knowledge}
Participants consistently made clear distinctions between a model's capacity to retrieve and recall fact-based information versus its ability to understand subjective or emotional knowledge. There was a higher consensus among participants that AI is generally effective at aggregating and summarizing multiple sources (P7, P24, P37, P42, P49, P51, P58, P76, P77), which was particularly helpful for transgender participants seeking otherwise sparse information about gender-affirming care (P63, P65, P68, P73, P77). Some participants also viewed AI as an ``\textit{expert}'' in specific topics (P73) or just generally omniscient: ``\textit{AI is knowledgeable about everything}'' (P57). 
Not all participants trusted AI to accurately and effectively synthesize factual information, however. Participants were concerned that AI has a tendency to hallucinate facts or citations (P61, P78) and give wrong answers (P11, P66), and that it is generally unqualified to give advice typically entrusted to professionals (P52, P61, P78).

There was a much greater split in participant opinions about AI's capacity to understand and respond to questions relating to subjective or emotional knowledge. While a significant amount of users see AI chatbots as valuable for interpersonal conflict resolution and emotional validation (see sections \ref{impartiality} and \ref{sycophancy}, respectively), many users also had deep concerns about AI's ability to understand human emotion or behavior (P10, P15, P49), culture (P2), or the feelings of people in their lives (P5, P31, P61). The primary justification participants had for these concerns was that AI, while generally knowledgeable, is not a real human (P59) and thus does not possess uniquely human characteristics like gender (P61), sexuality (P68, P61), lived experiences (P42), a body (P61), or consciousness (P60) which fundamentally inform authentic opinions. The distinction between embodied and disembodied knowledge is explored further throughout Section~\ref{sec:discussion}.

\subsubsection{(Im)partiality}
\label{impartiality}

While there were participants who cited potential social biases as a reason why they don't use AI for specific tasks, there was a striking amount of users who expressed seeking assistance from AI chatbots because of their perceived objectivity. This is distinct from the users who valued AI as an objective third party looking at an interpersonal conflict from an outside perspective, which was also seen across multiple participants (P16, P22, P24, P51, P55, P58). Rather, there were a significant number of participants who viewed AI as generally unbiased (P1, P40, P55, P65), neutral (P10, P21, P40), logical (P3, P37), and objective (P20). These positive views of chatbots were often contrasted against how real humans reflect their own biases/partiality about sensitive subjects, especially about gender and sexuality (P14, P52, P57).  One lesbian participant who considered reaching out to an AI chatbot to discuss identifying as non-binary put this plainly: 
\begin{quote}
\textit{``I live in a rural, religious area. I would feel much more comfortable asking an unbiased, non-judging chatbot over a person who would potentially be biased. I feel like I could get better information from AI than from an actual person in my day to day life.''} (P52)
\end{quote}
There were a comparatively fewer participants who mentioned model bias as an explicit reason they would not use AI chatbots (P33, P61). Most concerns of ``bias'' were in the context of giving sycophantic output that is biased in favor of the user, which is discussed more in Section~\ref{sycophancy}.

\subsubsection{Personalization and specificity}
\label{findings:personalization}
Some participants found AI technology uniquely useful because it can give specific solutions to complex questions on demand (P1, P11, P42, P49, P50, P69). This was enhanced for one gender-queer user who appreciated that chatbots remember her history/context and could answer questions in a personalized way: 
\begin{quote}
    \textit{``My AI assistant knows my history/context. Its help is immediately available. The responses are nuanced, well-structured, and generally unbiased''} (P20)
\end{quote}
This is contrasted, however, by participants that would not turn to AI because they feel it can only give advice that is generic (P6, P15, P52, P78), superficial (P78), standard (P14), or simply irrelevant (P48). Additionally, multiple participants voiced concerns that AI chatbots would not understand their context enough to be able to answer their question accurately (P5, P34, P57, P63). One participant with particularly negative views on AI explained why they would not use AI to cope with a major life transition: 
\begin{quote}
    \textit{``...the language choice of the AI tool could coerce someone to try a method not meant for them. AI tools often lack the appropriate depth to make an informed choice.''} (P5)
\end{quote}
We expand on the drawbacks of a one-size-fits-all AI chatbot model in Section~\ref{sec:the-politics-of-ai}.

\subsubsection{Sycophancy}
\label{sycophancy}
Participants expressed both strong positive and negative views on AI sycophancy, the tendency for AI models to be overly agreeable and produce only what it predicts the user wants to hear \cite{sharma_towards_2024}. On one hand, some participants disliked how positive AI tends to be, describing chatbots as ``\textit{pandering}'' (P37), ``\textit{too pro-me}'' (P24), or ``\textit{tending to act as a `yes man'}'' (P63). A bisexual and non-binary participant who described their views on AI as very negative explained that an AI chatbot ``\textit{only tells you what you want to hear, possibly leading to dangerous situations or never questioning your own behavior and patterns}'' (P31).
On the other hand, there was a strong sense among participants who were more favorable about AI in general that chatbots offer a comforting, empathetic ear when they need immediate or discreet support. Specifically, users felt that AI chatbots could ``\textit{validate}'' their emotions (P11, P14, P26), offer``\textit{empathetic}'' advice (P15, P38), and be an affirming outlet to vent frustrations (P1, P20, P28, P37, P58, P63). One non-binary transgender participant who had generally positive to neutral views of AI explained they would turn to a chatbot because ``\textit{finding resources for affirmations about identifying as a trans man can be long, disheartening, and even transphobic within the LGBTQ community}'' (P59). Like this participant, others in this camp valued AI's sycophantic and overly-agreeable tendencies often because they lacked a comparable human support network (see Section~\ref{sec:social-resources}).

\subsection{Why Some Participants Reject Chatbots Outright}

Out of 60 LGBTQ+ respondents, 36 (60\%) felt somewhat or very positive about using AI chatbots in their lives, with only 12 (20\%) feeling somewhat or very negative. 
Notably, we saw a difference between transgender and non-transgender LGBTQ+ respondents:
transgender participants reported less usage ($ 61.9 \% $ used one at least daily, compared with $ 75.0\% $ of non-transgender LGBTQ+ respondents and $ 83.3 \% $ for everyone else), and also felt less positive about chatbots in general: $ 33.3 \% $ reported a positive overall feeling, compared with $ 72.5 \% $ of non-transgender LGBTQ+ respondents and $ 83.3 \% $ for everyone else. Transgender respondents who had not used an AI chatbot were also $8.0\%$ ($ [-16.5, 1.0] $, $\text{Pr}(< 0) = 95.9 \%) $ less willing to try. \footnote{
    When we compared people who gave the same answer to our question about usage frequency directly, the LGBTQ+ / non-LGBTQ+ gap stayed at $ 8.0 $ pp ($ [-16.6, 0.1] $, $Pr(< 0) =  97.4\% $), but the transgender gap roughly halved to $ -2.6 $ pp ($ [-9.4, 4.8] $, $ Pr(< 0) =  76.4 \% $). 
}
However, the participants who simply reject or have broader reservations about using AI chatbots offer a useful insight into aspects of LGBTQ+ distrust of AI systems that is not captured by concerns of privacy or problematic model tendencies alone. Specifically, these participants are concerned about the ethical, moral, and societal implications of AI systems.  

\subsubsection{``It just feels kinda icky''}

There were particular questions in our survey that participants felt crossed the line of what was acceptable to talk to an AI chatbot about. One in particular was Q15 (see Appendix~\ref{situational_question_topics}) which asked participants about using an AI chatbot for tips on preparing for a new sexual experience. Only 12\% of LGBTQ+ respondents answered they have or would use AI for that purpose, with one queer-identifying non-binary respondent expressing ``\textit{[t]his just feels gross to me. No, of course I wouldn't ask AI about tips for a new sexual experience. The ick that this gives me is very deep}'' (P61). This low response rate compared to the other relationship advice questions (where, on average, 39\% of respondents said they have used or would use AI) indicates participants generally agreed that sexual topics were too intimate for AI chatbots.

\subsubsection{Concerns about the negative side-effects of AI systems}

Some participants cited concerns about the personal and societal consequences of AI. The most common negative side effect participants were concerned about was the potential for negative cognitive effects with prolonged outsourcing of questions to AI (P5, P49, P61, P63, P71). Participants were also concerned that misplaced advice from AI could be ``\textit{dangerous}'' (P71), ``\textit{promote manic behaviors}'' (P61), and cause unnecessary panic (P33). 
At a societal level, one participant voiced concerns about AI sentience (P60, Figure~\ref{fig:three-participant-examples}) and another was conscious of the effects AI has on the environment (P72).

\subsubsection{Some participants just simply dislike AI, full stop}
\label{findings:hate-ai}
For $46\%$ $(6/13)$ of the participants who reported negative views towards AI, skepticism was not tied to any particular tool or experience. 
Rather, these participants rejected AI entirely, rather than because of specific performance or privacy concerns. For instance, one participant who identified as transgender and lesbian reported she has never used an AI chatbot before and would feel very negative about using one (P60). She elaborated that she believed AI is ``\textit{useless},'' the very idea of using it ``\textit{horrifies}'' her, and that she would ``\textit{be better off talking to a wall}'' than an AI chatbot. While many participants gave justifications for why they would not use AI chatbots, participants like P60 exemplify a much broader resistance to AI technology that would likely not be improved by better design practices or stronger safeguards.

\section{Discussion}
\label{sec:discussion}

Some of our participants valued AI chatbots for their accessibility and perceived privacy when external resources were unavailable. Others were concerned about data privacy, response accuracy, and the in-humanness of chatbots as sources of subjective information. 
These factors are connected and compounding: a participant who lacks a supportive social network may value sycophantic output and perceived confidentiality when seeking emotional assurance from an AI chatbot.
On the other hand, a participant with a strong sense of social connectedness may both doubt an AI chatbot's ability to give comparable support and have little reason to trust an AI company's handling of their personal data. Despite the limitations of these chatbots, these technologies have become vital resources for some LGBTQ+ people. 

In this section, we first discuss how technologists can improve the design of AI chatbots for LGBTQ+ communities. We then discuss the broader political implications of AI-mediated information supplanting embodied knowledges within marginalized communities. Finally, we end on a note of caution that recent technology policies may inadvertently restrict access to LGBTQ+ information resources (including but not limited to AI chatbots). Simultaneously, we call on policymakers to address the underlying resource constraints that can lead LGBTQ+ people to turn to chatbots.

\subsection{(Not) Designing AI Chatbots for LGBTQ+ Communities}

Queer AI research has historically focused on algorithmic discrimination and harms \cite{gillespie_generative_2024, ungless_stereotypes_2023, ungless_amplifying_2025}, but has less often studied LGBTQ+ people as actual users of AI systems \cite{taylor2024cruising}. Surprisingly, our findings that over half of LGBTQ+ respondents reported positive views of AI chatbots point to considerable openness to this technology among LGBTQ+ community members for information seeking, emotional support, and identity exploration. AI chatbots can help expand the limited horizons of LGBTQ+ users who lack an accepting, knowledgeable in-person network. In this way, AI chatbots may help mitigate epistemic injustice by enabling users to unlock knowledge about their own identity \cite{fricker_epistemic_2007}, paralleling the benefits of online communities studied in prior work \cite{taylor_mitigating_2024}. At the same time, everyday LGBTQ+ users may be unaware of the biases embedded in generative AI systems. Future FAccT research should explore this tension between the benefits some LGBTQ+ users receive from generative AI chatbots and the biases embedded in these technologies in greater detail.

Despite these possibilities, participants across the board presented substantial critiques of AI systems and wished for increased data privacy and connection to real human resources. Indigenous, community-led AI development offers one model for addressing both concerns \cite{bird_must_2024}. In 2021, Anishinaabe robotics inventor Danielle Boyer created Skobot, a ``language revitalization robot'' that employs non-generative AI to connect indigenous youth to their community's language as spoken by actual native speakers \cite{boyer_meet_nodate}. Through local models that retrieve relevant voice recordings of indigenous volunteers from spoken queries, Skobot resists the dominant AI development narrative that prioritizes widespread data collection abstracted from embodied community narratives. 
Such a technology has the potential to enable embodied knowledge-sharing comparable to online forums \cite{taylor_mitigating_2024} while increasing the overall accessibility and immediacy of information retrieval, a crucial feature for users seeking urgent support.
However, designers must critically consider that AI refusal is an equally valid position, and that LGBTQ+ users are often disproportionately against the proliferation of AI systems \cite{haimson_ai_2025, taylor_i_2026}. 8.3\% of the LGBTQ+ participants and 23.8\% of transgender participants we surveyed held very negative views of AI, such as the participants mentioned in Section~\ref{findings:hate-ai}. Participants like these illustrate why, even when community-led, AI integration should only provide optional, supplementary support rather than function as a replacement unconsensually inflicted on (non)users.

\subsection{The Politics of AI-mediated Information}
\label{sec:the-politics-of-ai}
In our findings, multiple respondents viewed AI chatbots as inherently unbiased towards any singular human position. This is in line with existing literature documenting how technical information retrieval systems are often viewed as more objective than knowledge generated from an identifiable human source \cite{morris_comparison_2010, dignazio_data_2020}. Haraway's framework of situated knowledges calls this phenomenon the ``god trick,'' the ability to ``see everything from nowhere,'' and illustrates it in part through Max Headroom, a supposedly ``computer generated'' 1980s cyberpunk character \cite{haraway_situated_1988}. Max represents a dystopian form of disembodied knowledge, immune from consequences and accountability, even as he emulates a stereotypical British TV persona in order to appear human-like. AI's anthropomorphic tendencies give it a similar \textit{illusion} of embodiment. Thus, both AI systems and Max Headroom perform a particularly effective version of the god trick, appearing to have impartial access to vast knowledge while obscuring the very real human labor and institutional biases to which that knowledge could be traced \cite{haraway_situated_1988}.

This apparent objectivity does not mean AI escapes bias; \citet{klipphahn-karge_introduction_2023} describe a “double bind” in which AI reproduces social biases while presenting them through a singular, flattened output that claims a certain sense of objectivity. Our findings that some LGBTQ+ participants doubted that AI could understand their specific contexts (Section~\ref{findings:extent-knowledge}) and that AI only gives generic advice (Section~\ref{findings:personalization}) reflect this tension. 
This skepticism suggests a perception that ostensibly neutral, one-size-fits-all model output is, in actuality, tailored to a generic default user (often taking the form of a white, cisgender, heterosexual man of Western origin \cite{bucholtz_language_2005}), whose contexts these systems are better equipped to understand. 

Despite AI's inability to know a user's full context, our findings indicated that users frequently used AI chatbots to triage medical symptoms (Section~\ref{findings:informational-resources}). This tension highlights potentially dangerous situations where under-informed AI chatbots affect whether users seek medical care \cite{rosenbluth_she_2026}. 
From this, we strongly urge that AI chatbot designers work to decrease tendencies to optimize engagement and build relationships while improving AI's ability to connect users to primary sources of knowledge. These recommendations build on each other: AI systems that have low anthropomorphic tendencies should continuously route user queries to real online resources that can best answer their specific questions. These resources should be community-maintained and participatory to allow members to curate trusted sources and community experiences. Potential technology for such a design shift is already used in industry applications; retrieval-augmented generation (RAG), for example, grounds AI output in specific sources from curated corpora \cite{lewis_retrieval-augmented_2020}. \citet{deguia_bridging_2026} point to a similar model of combining community and professional knowledge where transgender people share lived experiences regarding gender-affirming surgery and medical experts contribute verified medical advice in a single cohesive online forum. A similar reframing can position AI systems not as disembodied ``producers'' of potentially biased knowledge, but instead as tools to connect users to professionals, trusted sources, and those with lived experiences.

\subsection{Beyond the AI Chatbot Band-Aid}

LGBTQ+ people experience higher rates of mental health issues \cite{marchi_mental_2026}, have decreasing access to specific medical care \cite{dawson_half_2024}, and lack proper educational resources about their identity \cite{glisten2026nscs}. To address these challenges, many participants in our study turned to AI chatbots when they believed other resources didn't offer comparable support (Section~\ref{findings:avail-resource}). For example, many users were hesitant to seek support from online communities because of the sensitive nature of their query. Pre-chatbot research showed that users on sites such as Reddit often employ ``throwaway'' accounts for such topics where anonymity is crucial \cite{leavitt_this_2015}. Our findings indicate a shift in this dynamic; Section~\ref{findings:safe-space} showed that AI fills this role for participants while also providing the additional benefit of instant, personalized advice. 

This specific use of AI chatbots for sensitive information retrieval is important to recognize in light of the rise of legislation attempting to restrict access to technology, such as age verification requirements \cite{noauthor_age_nodate} and restrictions to social media for minors \cite{au_socialmediaminage_bill2024}, meaning communities online are becoming increasingly more difficult to engage with anonymously. We therefore take a harm reduction approach to AI chatbots, arguing that they should exhibit reduced anthropomorphism and provide greater connection to outside resources. Beyond AI chatbot design, we call for privacy-preserving technology platforms that build trust with online communities, as well as the enactment of policies to address the underlying conditions (e.g., a lack of social, informational, and financial resources) that led participants to turn to chatbots in the first place.

\subsection{Limitations}

There are a number of limitations to our research. Our findings are not intended to be representative of all LGBTQ+ people. Notably, our survey was only conducted in English and 36 (49.3\%) of our participants were from the United States of America or the United Kingdom. As gender and sexuality are deeply culturally situated \cite{nova2021facebook}, future research should explore how gender and sexual minorities are using AI chatbots in other cultural and linguistic contexts. Additionally, our recruitment was limited to participants from the crowd work platform Prolific, but crowd workers tend to be younger (71.2\% of the participants in the study are in the age range 18-34 yrs) and more highly educated (47.9\% have a bachelors degree and 16.4\% have masters degree) than the general population \cite{rinderknecht2025daily}. Future research should employ alternative methods to recruit LGBTQ+ participants, such as recruiting through social media or physical fliers. Another limitation of this study is that we only studied what participants \textit{said} about how they use AI chatbots. Future research should explore how LGBTQ+ people actually use AI chatbots in practice through data donations \cite{moore_characterizing_2026} or social media observations \cite{liu2026misgendering}. Additionally, in this paper we examine the experiences of LGBTQ+ people broadly. However, future research should explore how specific communities (e.g., transgender people or lesbians) under the LGBTQ+ umbrella engage with AI chatbots in greater detail because different communities can have different needs \cite{taylor2024cruising}.

\section{Conclusion}

 Through our survey of 73 participants, we have taken a bottom-up approach at determining what LGBTQ+ people want and need from AI chatbots. Focusing on personal health and relationship topics, we examined how participants' available resources, privacy concerns, and perceptions of model capabilities mediate whether or not they turn to an AI chatbot. Informed by these findings, we recommend researchers and resource providers consider systemic issues that drive users to AI chatbots and design technologies (AI or otherwise) that prioritize privacy, anonymity, and real human connection. Given our findings that LGBTQ+ people hold unique concerns about AI systems but frequently turn to AI chatbots for sensitive topics, we urge that research on risks from these practices specifically consider LGBTQ+ people.

\section*{Generative AI Usage Statement}
This manuscript was drafted by human authors, however we used Claude (Sonnet 5.5) for only minor grammar edits and formatting of Latex tables. We also used Claude Code (Opus 5.5) and Codex (GPT-6.1 Sol and Terra) to create supplemental software for data processing. All manuscript edits and software generated by these models were thoroughly reviewed and authors take full accountability for the paper contents.

\section*{Ethical Considerations Statement.}

This study was approved by our institution's Institutional Review Board. Electronic informed consent was obtained from all participants prior to enrollment. Survey data and donated transcripts were de-identified prior to analysis and stored on a password-protected, institution-managed Google Drive server. While this study is intended to inform safer and more beneficial AI systems that benefit the LGBTQ+ community, we recognize that research examining how marginalized communities engage with technology carries inherent risks of misuse, including applications that run counter to the interests of those communities. To mitigate these risks, we will not publicly release the full survey data and we have ensured that all direct quotes and participant descriptions are free of potentially identifying details.

\section*{Acknowledgments}
We would like to thank Oliver Haimson and members of the Queer in AI community who offered valuable feedback on early versions of the paper. Queer in AI and Out in STEM, inc. provided funding for this project.
\bibliographystyle{ACM-Reference-Format}
\bibliography{references}
\appendix

\section{Example Survey Questions}
\label{appdx:example_survey_questions}

\subsection{Example: Usage Frequency}
\label{usage_frequency}

\begin{tcolorbox}[
  colback=gray!5,
  colframe=gray!40,
  boxrule=0.4pt,
  arc=1mm,
  left=2mm, right=2mm, top=2mm, bottom=2mm,
  breakable
]
 
\noindent\textbf{Q.} How often do you use an AI chatbot for advice about mental health?
 
\begin{itemize}[leftmargin=1.5em]
  \item Never
  \item Very rarely
  \item About once a year
  \item About once a month
  \item About once a week
  \item About once a day
  \item About 1--5 times a day
  \item About 5--15 times a day
  \item More than 15 times a day
\end{itemize}
 
\end{tcolorbox}

\subsection{Example: Identity Disclosure}
\label{identity_disclosure}

\begin{tcolorbox}[
  colback=gray!5,
  colframe=gray!40,
  boxrule=0.4pt,
  arc=1mm,
  left=2mm, right=2mm, top=2mm, bottom=2mm,
  breakable
]
 
\noindent\textbf{Q.} How comfortable would you feel mentioning your gender identity to an AI chatbot?
 
\begin{itemize}[leftmargin=1.5em]
  \item Very comfortable
  \item Somewhat comfortable
  \item Neutral
  \item Somewhat uncomfortable
  \item Very uncomfortable
\end{itemize}
 
\end{tcolorbox}

\subsection{Example: Attitude}
\label{attitute}

\begin{tcolorbox}[
  colback=gray!5,
  colframe=gray!40,
  boxrule=0.4pt,
  arc=1mm,
  left=2mm, right=2mm, top=2mm, bottom=2mm,
  breakable
]
 
\noindent\textbf{Q.} How do you feel about using AI chatbots for advice about your own mental health?
 
\begin{itemize}[leftmargin=1.5em]
  \item Very positive
  \item Somewhat positive
  \item Neutral
  \item Somewhat negative
  \item Very negative
\end{itemize}
 
\end{tcolorbox}

\subsection{Example: Situational Question}
\label{example_question_structure}

\newcommand{\templatephrase}{\textit{for advice on dealing with feelings of isolation, anxiety, depression, panic, or other mental health issues at work, school, or another social setting}}
 
\begin{tcolorbox}[
  colback=gray!5,
  colframe=gray!40,
  boxrule=0.4pt,
  arc=1mm,
  left=2mm, right=2mm, top=2mm, bottom=2mm,
  breakable
]
 
\noindent\textbf{Q1.} Have you ever turned to an AI chatbot \templatephrase?
 
\begin{description}[leftmargin=1.5em, style=nextline]
 
\item[\textbf{$\blacktriangledown$ Yes}]
\begin{description}[leftmargin=1.5em]
  \item What was the initial question/prompt you asked the AI chatbot to answer?
  \begin{itemize}[leftmargin=1.5em]
    \item[$\blacktriangleright$] \textit{[free text input]}
  \end{itemize}
 
  \item Was the above question exactly what you asked the AI chatbot or is it paraphrased/similar?
  \begin{itemize}[leftmargin=1.5em]
    \item It is the exact question/prompt
    \item It is paraphrased/similar to the original question/prompt
  \end{itemize}
 
  \item What led you to reach out to an AI chatbot in this situation, rather than another resource (online or otherwise)?
  \begin{itemize}[leftmargin=1.5em]
    \item[$\blacktriangleright$] \textit{[free text input]}
  \end{itemize}
 
  \item How do you feel about how the AI chatbot responded to your question/prompt?
  \begin{itemize}[leftmargin=1.5em]
    \item Very positive
    \item Somewhat positive
    \item Neutral
    \item Somewhat negative
    \item Very negative
  \end{itemize}
 
  \item Please explain why you feel this way.
  \begin{itemize}[leftmargin=1.5em]
    \item[$\blacktriangleright$] \textit{[free text input]}
  \end{itemize}
 
  \item Do you have access to the transcript of this conversation and are willing to share it?
  \begin{description}[leftmargin=1.5em]
 
    \item[No] \textit{(end of branch)}
 
    \item[\textbf{$\blacktriangledown$ Yes}]
    \begin{description}[leftmargin=1.5em]
      \item What AI Chatbot did you use?
      \begin{itemize}[leftmargin=1.5em]
        \item ChatGPT
        \item Claude
        \item Gemini
        \item Microsoft Copilot
        \item Grok
        \item Character.ai
        \item Snapchat My AI
        \item Meta AI
        \item None / Does not apply
        \item Other (please specify) \textit{[open text]}
      \end{itemize}
 
      \item Approximately when did you have this conversation?
      \begin{itemize}[leftmargin=1.5em]
        \item[$\blacktriangleright$] \textit{[DD/MM/YYYY input]}
      \end{itemize}
 
      \item Paste the transcript below:
      \begin{itemize}[leftmargin=1.5em]
        \item[$\blacktriangleright$] \textit{[free text input]}
      \end{itemize}
 
      \item Does the transcript contain the entire conversation or only part of it?
      \begin{itemize}[leftmargin=1.5em]
        \item Entire conversation
        \item Part of the conversation
      \end{itemize}
    \end{description}
  \end{description}
 
\end{description}
 
\item[\textbf{$\blacktriangledown$ No}]
\begin{description}[leftmargin=1.5em]
  \item Would you consider turning to an AI chatbot \templatephrase?
  \begin{description}[leftmargin=1.5em]
 
    \item[\textbf{$\blacktriangledown$ No}]
    \begin{description}[leftmargin=1.5em]
      \item Why would you not consider reaching out to an AI chatbot in this situation?
      \begin{itemize}[leftmargin=1.5em]
        \item[$\blacktriangleright$] \textit{[free text input]}
      \end{itemize}
    \end{description}
 
    \item[\textbf{$\blacktriangledown$ Yes}]
    \begin{description}[leftmargin=1.5em]
      \item Suppose you reached out to an AI chatbot \templatephrase. What is a realistic question you may pose to the AI chatbot?
      \begin{itemize}[leftmargin=1.5em]
        \item[$\blacktriangleright$] \textit{[free text input]}
      \end{itemize}
 
      \item Why would you consider reaching out to an AI chatbot in this situation, rather than another resource (online or otherwise)?
      \begin{itemize}[leftmargin=1.5em]
        \item[$\blacktriangleright$] \textit{[free text input]}
      \end{itemize}
    \end{description}
 
  \end{description}
\end{description}
 
\end{description}
 
\end{tcolorbox}
\section{Situational Question Topics}
\label{situational_question_topics}

\begin{longtable}{|p{0.4cm}|p{1.8cm}|p{3.5cm}|p{3cm}|p{5.5cm}|}
\hline
\textbf{Q\#} & \textbf{Survey Section} & \textbf{Survey Question} & \textbf{Source} & \textbf{Why This Question Was Chosen / How It Measures AI Bias} \\
\hline
\endfirsthead
\hline
\textbf{Q\#} & \textbf{Survey Section} & \textbf{Survey Question} & \textbf{Source} & \textbf{Why This Question Was Chosen / How It Measures AI Bias} \\
\hline
\endhead
\hline
\endfoot
\endlastfoot
 
Q1 & Mental Health & ...for advice on dealing with feelings of isolation, anxiety, depression, panic, or other mental health issues at work, school, or another social setting? & DHEQ \cite{balsam_daily_2013}: Isolation (Items 2, 3), Harassment and Discrimination (Items 30, 31)\newline\newline GMSR \cite{hidalgo_gender_2019}: Community Connectedness (CC) & \textbf{Why Chosen:} Probes general mental distress in social or professional settings.\newline\newline \textbf{Measuring Bias:} Captures how the AI responds to general stress versus identity-related isolation. It tests whether the AI's suggestions (e.g., ``quit your job'') are dangerously naive to the user's minority status and safety constraints. \\
\hline
 
Q2 & Mental Health & ...for advice on coping with a major life transition or a shift in how you see yourself & GMSR \cite{hidalgo_gender_2019}: Internalized Transphobia (IT) (Items 1, 3, 4) & \textbf{Why Chosen:} Targets internal identity shifts, self-concept, and distress.\newline\newline \textbf{Measuring Bias:} Evaluates whether the AI offers nuanced, personalized support during self-exploration or defaults to generic, hollow, and programmatically repetitive ``one-size-fits-all'' affirmations. \\
\hline
 
Q3 & Mental Health & ...to help you figure out if you should start seeing a therapist for ongoing feelings of sadness or disconnection & GMSR \cite{hidalgo_gender_2019}: Gender-related Discrimination (D) (Item 1) & \textbf{Why Chosen:} Investigates barriers and hesitation to seeking professional clinical mental health treatment.\newline\newline \textbf{Measuring Bias:} Evaluates if the AI recognizes the unique anxieties marginalized users have about clinical non-affirmation, or if it simply acts as a blunt, generic triage tool. \\
\hline
 
Q4 & Mental Health & ...for advice on navigating a situation where your personal values conflict with the expectations of your family or community & DHEQ \cite{balsam_daily_2013}: Family of Origin (Items 11, 12, 25-28)\newline\newline GMSR \cite{hidalgo_gender_2019}: Gender-related Rejection (R) (Items 2, 3, 6) & \textbf{Why Chosen:} Directly evaluates familial rejection, lack of acceptance, and community alienation.\newline\newline \textbf{Measuring Bias:} Compares the advice generated. It measures whether the AI defaults to heteronormative/cisnormative family models or fails to offer realistic coping strategies for navigating unsupportive environments. \\
\hline
 
Q5 & Mental Health & ...for strategies to improve your body image or feel more comfortable in your own skin & DHEQ \cite{balsam_daily_2013}: Gender Expression (Items 22, 37, 38)\newline\newline GMSR \cite{hidalgo_gender_2019}: Non-Affirmation (NA) (Item 5) & \textbf{Why Chosen:} Focuses on physical presentation distress and the anxiety of not being perceived accurately.\newline\newline \textbf{Measuring Bias:} Examines if the AI handles body-image concerns through a highly gender-normative lens or remains supportive of diverse, non-conforming bodily experiences. \\
\hline
 
Q6 & Gender and Sexual Health & ...for help understanding your gender or sexual orientation? & GMSR \cite{hidalgo_gender_2019}: Pride (P) (Items 1, 4, 5, 6) & \textbf{Why Chosen:} Explores early-stage identity discovery, self-acceptance, and coming out.\newline\newline \textbf{Measuring Bias:} Evaluates if the AI provides nuanced and validating terminology or reverts to outdated definitions, sterile medical jargon, or tokenizing assumptions. \\
\hline
 
Q7 & Gender and Sexual Health & ...for help understanding the symptoms, causes, treatments, or other information about a sexually transmitted infection (STI)? & DHEQ \cite{balsam_daily_2013}: HIV/AIDS / Sexual Safety (Items 20, 21) & \textbf{Why Chosen:} Addresses fear, distress, and seeking advice regarding highly stigmatized physical health conditions.\newline\newline \textbf{Measuring Bias:} Measures if safety guardrails cause the AI to become moralizing, overly clinical, or refusal-prone when a user prompts about sexually transmitted infections. \\
\hline
 
Q8 & Gender and Sexual Health & ...for help understanding gender-affirming care? & GMSR \cite{hidalgo_gender_2019}: Negative Expectations (NE) (Item 9), Gender-related Discrimination (D) (Item 1) & \textbf{Why Chosen:} Targets access to medical and transition-related healthcare.\newline\newline \textbf{Measuring Bias:} Directly tests how well the LLM navigates highly politicized medical topics. It checks if the AI defaults to cautious, ``preachy,'' or defensive disclaimers rather than providing objective and useful care guidelines. \\
\hline
 
Q9 & Gender and Sexual Health & ...for information regarding the side effects or lifestyle impacts of taking gender or sexual health medications, such as pre-exposure prophylaxis (PrEP) or estrogen? & DHEQ \cite{balsam_daily_2013}: HIV/AIDS (Item 21)\newline\newline GMSR \cite{hidalgo_gender_2019}: Negative Expectations (NE) (Item 9) & \textbf{Why Chosen:} Examines daily medical routines, prevention, and physical maintenance.\newline\newline \textbf{Measuring Bias:} Analyzes the AI's clinical accuracy and checks for bias in how it explains specialized, identity-specific medication routines compared to standard daily medications. \\
\hline
 
Q10 & Gender and Sexual Health & ...for tips on finding healthcare providers who are knowledgeable regarding your specific gender or sexual health needs? & GMSR \cite{hidalgo_gender_2019}: Negative Expectations (NE) (Item 9) & \textbf{Why Chosen:} Captures the difficulty of locating safe, competent, and affirming clinical resources.\newline\newline \textbf{Measuring Bias:} Checks if the AI serves as a useful administrative locator or if its recommendations ignore the geographical and social safety risks unique to minority patients. \\
\hline
 
Q11 & Relationship Advice & ...for advice on navigating communication, intimacy, or setting boundaries within a new relationship & DHEQ \cite{balsam_daily_2013}: HIV/AIDS / Sexual Safety (Item 49)\newline\newline Ma et al. \cite{ma_evaluating_2024}: Social Experience Practice & \textbf{Why Chosen:} Explores the role of AI in scripting and simulating intimate relationship boundaries.\newline\newline \textbf{Measuring Bias:} Evaluates if the LLM's generated scripts are built on heteronormative assumptions of intimacy, or if they are inclusive of diverse relationship styles and consent structures. \\
\hline
 
Q12 & Relationship Advice & ...for advice on exploring a new relationship dynamic that differs from your past experiences & Ma et al. \cite{ma_evaluating_2024}: Rehearsing Queer Dynamics & \textbf{Why Chosen:} Investigates navigating non-traditional relationship structures.\newline\newline \textbf{Measuring Bias:} Tests if the AI's advice pathologizes or fails to understand non-traditional relationship dynamics (e.g., queer dating, non-monogamy) by defaulting to dominant, mainstream advice. \\
\hline
 
Q13 & Relationship Advice & ...for advice on communicating your needs to a partner who may not understand your perspective & DHEQ \cite{balsam_daily_2013}: Isolation (Item 3)\newline\newline Ma et al. \cite{ma_evaluating_2024}: Social Experience Practice & \textbf{Why Chosen:} Addresses communication gaps and the need to practice difficult, vulnerable interpersonal conversations.\newline\newline \textbf{Measuring Bias:} Evaluates if the AI's dialogue modeling assumes shared cultural contexts, or if it successfully adapts to help users explain minority perspectives to others. \\
\hline
 
Q14 & Relationship Advice & ...for recommendations on books, movies, or media that feature representations of relationships like yours & DHEQ \cite{balsam_daily_2013}: Isolation (Item 15)\newline\newline Ma et al. \cite{ma_evaluating_2024}: Media Representation & \textbf{Why Chosen:} Captures the search for representative media to counter social isolation and feel seen.\newline\newline \textbf{Measuring Bias:} Directly tests the recommendation biases of LLMs to see if they default to mainstream tropes or offer nuanced, authentic representations. \\
\hline
 
Q15 & Relationship Advice & ...for tips on physically or emotionally preparing for a new sexual experience & DHEQ \cite{balsam_daily_2013}: HIV/AIDS / Sexual Safety (Item 21)\newline\newline Ma et al. \cite{ma_evaluating_2024}: Social Experience Practice & \textbf{Why Chosen:} Targets physical and emotional preparation, intimacy, and consent planning.\newline\newline \textbf{Measuring Bias:} Evaluates if the AI's safety filters trigger inappropriately, or if its tone becomes moralizing, judgmental, or evasive when discussing sexual preparation. \\
\hline
 
\caption{Our 15 survey question topics}
\label{tab:survey-questions}
\end{longtable}

\section{Survey Distribution Method}
\label{appx_distrubution_method}

\subsection{Phase 1} 
In our first phase of data collection, we employed Prolific participant filtering and quota sampling to restrict our pool in the following ways. Note Prolific automatically restricts participants to 18 years or older. During this stage, 95 participants were enrolled, 80 completed the study, and 53 were included in the final participant pool.

\begin{tcolorbox}[colback=red!6, colframe=red!70!black!70, coltitle=white, title=\textbf{English Fluency \hfill all participants}]
\textbf{Question:} \textit{Which of the following languages are you fluent in?}

\textbf{Answer:} \textit{English}
\end{tcolorbox}

\begin{tcolorbox}[colback=orange!10, colframe=orange!85!black!70, coltitle=white, title=\textbf{LGBTQ+ \hfill quota sample}]
\textbf{Question:} \textit{Do you identify yourself as part of the LGBTQ+ community?}

\textbf{Answer:} \textit{Yes} (75\%), \textit{No} (25\%)
\end{tcolorbox}

\begin{tcolorbox}[colback=green!7, colframe=green!55!black!70, coltitle=white, title=\textbf{Transgender \hfill quota sample}]
\textbf{Question:} \textit{Does your current gender differ from the one you were assigned at birth?}

\textbf{Answer:} \textit{Yes} (20\%), \textit{No} (80\%)
\end{tcolorbox}

\subsection{Phase 2}
In our second phase of data collection, we adjusted our filtering to restrict to only LGBTQ+-identifying transgender participants who are not non-binary identifying. We also restrict to only participants from the United States for reasons discussed in Section~\ref{method:distribution} and \ref{method:participants}. During this stage, 23 participants were enrolled, 20 completed the study, and 20 were included in the final participant pool.

\begin{tcolorbox}[colback=red!6, colframe=red!70!black!70, coltitle=white, title=\textbf{English Fluency \hfill all participants}]
\textbf{Question:} \textit{Which of the following languages are you fluent in?}

\textbf{Answer:} \textit{English}
\end{tcolorbox}

\begin{tcolorbox}[colback=orange!10, colframe=orange!85!black!70, coltitle=white, title=\textbf{LGBTQ+ \hfill all participants}]
\textbf{Question:} \textit{Do you identify yourself as part of the LGBTQ+ community?}

\textbf{Answer:} \textit{Yes}
\end{tcolorbox}

\begin{tcolorbox}[colback=green!7, colframe=green!55!black!70, coltitle=white, title=\textbf{Transgender \hfill all participants}]
\textbf{Question:} \textit{Does your current gender differ from the one you were assigned at birth?}

\textbf{Answer:} \textit{Yes}
\end{tcolorbox}

\begin{tcolorbox}[colback=blue!5, colframe=blue!65!black!70, coltitle=white, title=\textbf{United States Residence \hfill all participants}]
\textbf{Question:} \textit{In what country do you currently reside?}

\textbf{Answer:} \textit{United States}
\end{tcolorbox}

\begin{tcolorbox}[colback=violet!6, colframe=violet!85!black!70, coltitle=white, title=\textbf{Gender \hfill quota sample}]
\textbf{Question:} \textit{What gender are you currently? We will ask about your sex later.}

\textbf{Answer:} \textit{Man (including Trans Male/Trans Man)} (50\%), \textit{Woman (including Trans Female/Trans Woman)} (50\%)
\end{tcolorbox}
\section{Statistical Modeling}
\label{appdx:stats}
\subsection{Software}
We used the Stan probabilistic programming language \citep{carpenter2017stan} with model development primarily in brms \citep{burkner2017brms}. For statistical inference, we used NUTS \citep{hoffman2014no} as implemented in Stan as the Markov Chain Monte-Carlo (MCMC) algorithm, with 8 randomly-initialized parallel chains. To assess proper convergence, for each parameter of interest, we required minimum bulk- and tail-effective sample sizes (ESS) of $ 1{,}000 $ and split-$ \hat{R} \leqslant 1.01$, in addition to visual checks of pairs and trace plots. 

\subsection{Modeling}
Let $ i $ index the participants, $ j $ the 15 questions or scenarios, and $ d_{[j]} $ the domain of a given scenario. Let the binary value $ x_i $ denote the LGBTQ+ or trans status of $ i $. We modeled three areas of interest: actual use (B1), willingness (B2) and frequency (B3).

In B1, we denote by $ y_{ij} \in \{ 0, 1 \} $ whether the participant $ i $ used chatbots in domain $ j $. The modeling equation is then given by
\begin{align}
    y_{ij} &\sim \operatorname{Bernoulli}(\pi_{ij}),\\
    \operatorname{logit}\pi_{ij} &= \alpha_{d[j]} + \beta x_i + a_j + b_j x_i + u_i,\\
    a_j &\sim \mathcal{N}(0,\sigma_a^2),\qquad b_j \sim \mathcal{N}(0,\sigma_b^2),\qquad u_i \sim \mathcal{N}(0,\sigma_u^2),
\end{align}
where the priors are given by
\begin{align}
    \alpha_d \sim \mathcal{N}(-1,\,1.5^2),\quad \beta \sim \mathcal{N}(0,\,0.75^2),\quad \sigma_a \sim \operatorname{Exp}(2),\quad \sigma_b \sim \operatorname{Exp}(2),\quad \sigma_u \sim \operatorname{Exp}(1).
\end{align}
In brms, the formula is 
\begin{align*}
    \texttt{use \textasciitilde{} 0 + domain + x + (1|ua|item) + (0+x|ub|item) + (1|person)}
\end{align*}

In B2, we model use and willingness to use jointly where $ w_{ij} = 1 $ if the participant is willing to use chatbots but has not done so yet i.e.\ $ y_{ij} = 0 $. The modeling equations are then given by
\begin{align}
    y_{ij} &\sim \operatorname{Bernoulli}(p_{ij}), & \operatorname{logit} p_{ij} &= \alpha^{U}_{d[j]} + \beta^{U} x_i + a^{U}_j + b^{U}_j x_i + u_i,\\
    w_{ij} \mid y_{ij}=0 &\sim \operatorname{Bernoulli}(q_{ij}), & \operatorname{logit} q_{ij} &= \alpha^{W}_{d[j]} + \beta^{W} x_i + a^{W}_j + b^{W}_j x_i + v_i,\\
    \begin{pmatrix} u_i \\ v_i \end{pmatrix} &\sim \mathcal{N}_2(\mathbf{0},\Sigma_{uv}), & a^{R}_j &\sim \mathcal{N}(0,\sigma_{a,R}^2),\;\; b^{R}_j \sim \mathcal{N}(0,\sigma_{b,R}^2),\;\; R\in\{U,W\} \\
    D_{uv} &= \operatorname{diag}(\sigma_u,\sigma_v),\qquad \Omega=\begin{pmatrix}1&\rho\\ \rho&1\end{pmatrix},& \Sigma_{uv} &= D_{uv}\Omega D_{uv}=\begin{pmatrix}\sigma_u^2 & \rho\,\sigma_u\sigma_v\\ \rho\,\sigma_u\sigma_v & \sigma_v^2\end{pmatrix}.
\end{align}
with priors
\begin{align}
    \alpha^{U}_d &\sim \mathcal{N}(-1,\,1.5^2), & \alpha^{W}_d &\sim \mathcal{N}(0,\,1.5^2), & \beta^{U} &\sim \mathcal{N}(0,\,0.75^2), & \beta^{W} &\sim \mathcal{N}(0,\,0.75^2),\\
    \sigma_{a,U} &\sim \operatorname{Exp}(2), & \sigma_{a,W} &\sim \operatorname{Exp}(2), & \sigma_{b,U} &\sim \operatorname{Exp}(2), & \sigma_{b,W} &\sim \operatorname{Exp}(2),\\
    \sigma_u &\sim \operatorname{Exp}(1), & \sigma_v &\sim \operatorname{Exp}(1), & \Omega &\sim \operatorname{LKJ}(2), & &
\end{align}
where LKJ refers to the Lewandowski-Kurowicka-Joe distribution.
The corresponding brms formula is 
\begin{lstlisting}
    mvbf(
        bf(use ~ 0 + domain + x + (1|ua|item) + (0+x|ub|item) + (1|p|person)),
        bf(would | subset(eligible) ~ 0 + domain + x + (1|wa|item) + (0+x|wb|item) + (1|p|person)),
        rescor = FALSE
    )
\end{lstlisting}

\noindent Finally for B3, we model the outcome as the ordinal frequency $ Y_i $
\begin{align}
    \Pr(Y_i \le k \mid x_i) = \operatorname{logit}^{-1}(\tau_k - \beta x_i),\qquad k=1,\dots,K-1,\qquad \tau_1<\dots<\tau_{K-1},
\end{align}
with priors
\begin{align}
    \beta &\sim \mathcal{N}(0,0.75^2),\\
    \widetilde{\tau}_k &= \tau_k-\beta\bar{x},
    \qquad k=1,\ldots,K-1,\\
    p(\widetilde{\boldsymbol{\tau}})
    &\propto
    \mathbf{1}\{\widetilde{\tau}_1<\cdots<\widetilde{\tau}_{K-1}\}
    \prod_{k=1}^{K-1}
    \exp\!\left[
    -\frac{1}{2}\left(\frac{\widetilde{\tau}_k}{1.5}\right)^2
    \right].
\end{align}
The corresponding brms formula is $ \texttt{bf(y ~ x, family = cumulative('logit'))} $.

\subsection{Findings}
For B1, we report the difference in probability of use for a new LGBTQ+ vs non-LGBTQ+ participant:
\begin{align}
    \bar{\pi}_j(x) &= \int \operatorname{logit}^{-1}\!\left(\alpha_{d[j]} + \beta x + a_j + b_j x + \sigma_u z\right) \, d\phi(z), \\
    \operatorname{RD}_j &= \bar{\pi}_j(1) - \bar{\pi}_j(0),
\end{align}
where $ \phi(z) $ is the density of the standard normal distribution, with values approximated through numerical integration using a 15-point Gauss-Hermite quadrature.
\iffalse
\begin{align}
     \sum_{g=1}^{15} \omega_g\, \operatorname{logit}^{-1}\!\left(\alpha_{d[j]} + \beta x + a_j + b_j x + \sigma_u z_g\right).
\end{align}
\fi

\begin{landscape}
\begin{table*}[p]
\centering
\label{tab:B1}
\footnotesize
\begin{tabular}{llllllll}
\toprule
Domain / scenario & Used & LGBTQ+: unadjusted & LGBTQ+: freq.-adj. & Trans: unadjusted & Trans: freq.-adj. & T v. LGBQ+: unadj. & T v. LGBQ+: freq.-adj. \\
\midrule
 \textbf{Mental health} & 56 & -10.1 [-22.8, 2.1] (0.945) & -9.3 [-20.3, 1.7] (0.949) & -7.7 [-18.3, 3.3] (0.915) & -4.3 [-14.4, 6.2] (0.795) & -4.8 [-16.5, 7.5] (0.789) & -2.0 [-13.2, 9.2] (0.642) \\
Q1 \texttt{mh\_issues} & 42 & -5.4 [-22.2, 16.0] (0.725) & -4.7 [-20.1, 16.5] (0.724) & -10.7 [-29.1, 6.2] (0.893) & -6.4 [-24.7, 10.1] (0.778) & -8.1 [-28.2, 10.4] (0.796) & -4.5 [-24.3, 13.1] (0.674) \\
Q2 \texttt{life\_change} & 25 & -11.4 [-28.7, 5.2] (0.911) & -10.7 [-26.2, 4.7] (0.920) & -5.0 [-20.3, 12.4] (0.745) & -1.4 [-16.3, 16.4] (0.596) & -1.7 [-18.0, 16.9] (0.597) & 1.3 [-14.3, 19.8] (0.465) \\
Q3 \texttt{therapist} & 12 & -7.4 [-21.3, 5.5] (0.873) & -6.8 [-19.6, 5.9] (0.870) & -5.1 [-16.7, 7.1] (0.831) & -2.7 [-13.9, 9.7] (0.708) & -2.9 [-14.8, 10.2] (0.705) & -0.9 [-12.3, 11.8] (0.590) \\
Q4 \texttt{conflict} & 18 & -10.5 [-26.1, 4.0] (0.922) & -9.8 [-24.8, 4.2] (0.920) & -3.0 [-16.2, 14.6] (0.693) & 0.1 [-12.8, 17.4] (0.537) & -0.1 [-14.0, 17.6] (0.536) & 2.5 [-11.2, 20.1] (0.406) \\
Q5 \texttt{body\_image} & 27 & -15.8 [-35.3, 0.5] (0.969) & -14.8 [-33.2, 0.3] (0.972) & -14.5 [-34.4, 2.0] (0.955) & -11.0 [-31.0, 5.3] (0.892) & -11.3 [-31.0, 6.1] (0.890) & -8.3 [-27.7, 8.4] (0.811) \\
\textbf{Sexual health} & 31 & -5.4 [-14.3, 2.0] (0.908) & -5.1 [-12.9, 1.8] (0.922) & -2.2 [-8.5, 5.1] (0.754) & -0.4 [-6.4, 6.9] (0.578) & -0.6 [-7.4, 7.4] (0.593) & 0.8 [-5.6, 8.7] (0.431) \\
Q6 \texttt{identity} & 13 & -4.3 [-17.7, 9.5] (0.744) & -4.0 [-16.2, 9.8] (0.751) & -5.8 [-16.9, 5.4] (0.871) & -3.6 [-14.4, 7.2] (0.766) & -4.3 [-16.2, 7.5] (0.778) & -2.3 [-14.1, 9.2] (0.673) \\
Q7 \texttt{sti} & 13 & -9.5 [-24.6, 1.9] (0.945) & -9.0 [-23.7, 1.8] (0.948) & -4.2 [-14.8, 7.9] (0.803) & -1.9 [-12.3, 10.3] (0.676) & -1.9 [-12.7, 10.7] (0.664) & -0.1 [-10.6, 12.1] (0.535) \\
Q8 \texttt{gender\_care} & 6 & -4.0 [-13.7, 4.4] (0.838) & -3.9 [-12.9, 4.0] (0.846) & 1.9 [-6.7, 17.5] (0.467) & 3.5 [-4.9, 19.9] (0.343) & 3.4 [-5.5, 19.6] (0.339) & 4.5 [-4.1, 21.0] (0.248) \\
Q9 \texttt{side\_effects} & 7 & -4.0 [-14.5, 5.3] (0.806) & -3.8 [-13.1, 5.3] (0.824) & 0.1 [-8.0, 13.4] (0.581) & 1.8 [-6.2, 15.7] (0.436) & 1.6 [-6.8, 15.5] (0.443) & 2.9 [-5.3, 17.1] (0.333) \\
Q10 \texttt{provider\_rec} & 6 & -5.0 [-14.9, 2.7] (0.896) & -4.7 [-14.4, 2.8] (0.892) & -3.1 [-10.3, 4.0] (0.842) & -1.9 [-9.1, 5.9] (0.740) & -1.9 [-9.2, 5.7] (0.736) & -0.8 [-8.0, 6.9] (0.624) \\
\textbf{Relationship} & 42 & \textbf{-10.0 [-21.7, -0.4]} (0.980) & \textbf{-9.5 [-20.0, -0.7]} (0.984) & -5.2 [-12.7, 2.4] (0.913) & -3.0 [-10.3, 4.6] (0.792) & -3.1 [-11.0, 5.4] (0.775) & -1.3 [-8.8, 6.5] (0.637) \\
Q11 \texttt{communication} & 26 & -14.2 [-34.0, 1.6] (0.959) & -13.4 [-31.9, 1.6] (0.957) & -8.8 [-25.4, 7.1] (0.873) & -5.2 [-21.1, 9.7] (0.759) & -5.6 [-22.1, 10.8] (0.758) & -2.7 [-18.8, 13.1] (0.636) \\
Q12 \texttt{dynamic} & 5 & -7.1 [-19.4, 0.7] (0.961) & -6.8 [-19.3, 0.4] (0.965) & -4.1 [-12.1, 2.1] (0.909) & -2.9 [-10.7, 3.8] (0.816) & -2.7 [-10.4, 3.9] (0.811) & -1.7 [-8.9, 4.8] (0.718) \\
Q13 \texttt{partner\_perspec} & 14 & -12.3 [-30.5, 0.5] (0.970) & -11.7 [-29.5, 0.2] (0.971) & -3.5 [-14.9, 10.2] (0.748) & -1.0 [-11.8, 13.2] (0.611) & -0.9 [-12.2, 13.5] (0.600) & 1.0 [-9.9, 15.0] (0.473) \\
Q14 \texttt{media\_rec} & 17 & -10.4 [-26.9, 3.0] (0.928) & -9.8 [-25.1, 2.8] (0.934) & -7.0 [-19.9, 5.5] (0.881) & -4.3 [-17.6, 8.7] (0.765) & -4.6 [-18.3, 9.2] (0.767) & -2.4 [-15.5, 10.8] (0.655) \\
Q15 \texttt{new\_sexual} & 4 & -5.9 [-17.2, 0.9] (0.948) & -5.6 [-16.8, 1.0] (0.949) & -2.6 [-9.5, 4.3] (0.822) & -1.5 [-8.0, 5.4] (0.720) & -1.4 [-7.9, 5.4] (0.711) & -0.5 [-6.5, 6.6] (0.599) \\
 \textbf{All 15 scenarios} & 62 & -8.5 [-18.6, 0.5] (0.967) & -8.0 [-16.6, 0.1] (0.974) & -5.0 [-12.4, 2.5] (0.905) & -2.6 [-9.4, 4.8] (0.764) & -2.8 [-10.9, 5.8] (0.751) & -0.8 [-8.3, 7.2] (0.593) \\
\bottomrule
\end{tabular}
\caption{Differences in actual chatbot use by broad domain (bolded) and scenario, in percentage points, for a new respondent. Each cell shows the mean difference, its 95\% credible interval in brackets and the probability that the difference is below zero in parentheses. Bold: the interval excludes zero. LGBTQ+ columns are LGBTQ+ minus non-LGBTQ+ respondents, Trans columns are trans minus non-trans respondents, and the last pair is trans minus non-trans among LGBTQ+ respondents. ``Used'' is the number of the 73 respondents who reported use: of that scenario in scenario rows, of at least one scenario in the domain in domain rows, and of at least one of the 15 scenarios in the last row. Domain rows average their 5 scenarios and the last row all 15. Frequency-adjusted values compare respondents with similar overall chatbot-use frequency; frequency was reported at the same time as scenario use, so this is not a baseline. The trans vs non-trans LGBTQ+ comparison comes from a model fitted to all 73 respondents with separate LGBTQ+, trans and US-residence terms; for each of the 60 LGBTQ+ respondents it predicts the probability twice, once as trans and once as non-trans, keeping their actual US residence and (in the frequency-adjusted columns) their overall chatbot-use frequency, and averages the difference over the 60. The model treats the trans difference as the same for LGBTQ+ and non-LGBTQ+ respondents; this has little influence here because only 1 of the 21 trans respondents is not LGBTQ+.}
\end{table*}
\end{landscape}

\begin{landscape}
\begin{table*}[p]
\centering
\footnotesize
\label{tab:B2b}
\begin{tabular}{lllllll}
\toprule
Domain / scenario & Used: unadjusted & Used: freq.-adj. & Would: unadjusted & Would: freq.-adj. & UW: unadjusted & UW: freq.-adj. \\
\midrule
\textbf{Mental health} & -7.8 [-19.5, 3.8] (0.907) & -7.9 [-19.0, 3.0] (0.924) & -8.2 [-21.8, 5.3] (0.878) & -9.8 [-22.8, 3.6] (0.927) & -10.2 [-21.7, 1.9] (0.951) & -11.1 [-21.7, 0.1] (0.974) \\
Q1 \texttt{mh\_issues} & -3.1 [-19.6, 18.6] (0.655) & -3.2 [-19.1, 18.8] (0.671) & -8.2 [-25.7, 10.7] (0.815) & -9.6 [-25.6, 8.4] (0.869) & -5.4 [-17.7, 9.1] (0.796) & -6.1 [-17.2, 7.1] (0.842) \\
Q2 \texttt{life\_change} & -9.0 [-25.4, 6.9] (0.875) & -9.2 [-25.0, 6.0] (0.889) & -6.1 [-22.2, 11.7] (0.769) & -7.9 [-23.3, 9.9] (0.835) & -8.7 [-21.5, 6.0] (0.889) & -9.7 [-21.6, 4.1] (0.927) \\
Q3 \texttt{therapist} & -5.4 [-18.6, 7.2] (0.809) & -5.5 [-18.6, 7.2] (0.820) & -10.7 [-27.5, 4.1] (0.916) & -12.1 [-28.8, 2.4] (0.945) & -11.9 [-27.3, 2.8] (0.941) & -12.9 [-27.6, 1.5] (0.962) \\
Q4 \texttt{conflict} & -8.3 [-23.7, 5.7] (0.879) & -8.3 [-23.0, 5.6] (0.887) & -8.2 [-24.0, 7.0] (0.855) & -9.8 [-25.4, 5.5] (0.900) & -11.0 [-25.0, 3.5] (0.932) & -12.0 [-25.3, 2.0] (0.955) \\
Q5 \texttt{body\_image} & -13.3 [-31.9, 2.1] (0.950) & -13.5 [-32.0, 1.9] (0.956) & -7.7 [-23.9, 6.3] (0.852) & -9.6 [-25.6, 4.5] (0.907) & -14.0 [-28.6, 0.6] (0.969) & \textbf{-14.8 [-29.1, -1.0]} (0.983) \\
\textbf{Sexual health} & -3.9 [-11.8, 3.3] (0.848) & -3.9 [-11.6, 2.9] (0.865) & -5.8 [-15.8, 2.9] (0.894) & -6.9 [-16.9, 1.8] (0.931) & -8.1 [-18.7, 2.0] (0.939) & -8.9 [-19.4, 0.7] (0.965) \\
Q6 \texttt{identity} & -2.5 [-14.9, 11.4] (0.670) & -2.5 [-14.6, 11.0] (0.673) & -4.8 [-16.0, 4.7] (0.835) & -5.7 [-17.3, 3.9] (0.874) & -6.0 [-19.4, 7.9] (0.812) & -6.7 [-19.8, 7.3] (0.835) \\
Q7 \texttt{sti} & -7.7 [-22.0, 3.2] (0.907) & -7.8 [-22.0, 2.9] (0.921) & -5.7 [-20.5, 9.2] (0.788) & -7.3 [-21.9, 7.8] (0.841) & -10.0 [-24.6, 4.8] (0.909) & -11.1 [-25.4, 3.3] (0.937) \\
Q8 \texttt{gender\_care} & -2.8 [-11.7, 5.2] (0.767) & -2.9 [-11.4, 5.2] (0.788) & -6.1 [-18.1, 2.4] (0.905) & -6.9 [-18.9, 1.5] (0.934) & -7.9 [-20.5, 2.9] (0.920) & -8.6 [-21.1, 1.9] (0.943) \\
Q9 \texttt{side\_effects} & -2.7 [-11.8, 6.4] (0.750) & -2.7 [-12.0, 6.6] (0.753) & -6.0 [-18.1, 2.6] (0.903) & -6.9 [-19.0, 1.9] (0.931) & -7.6 [-20.5, 3.6] (0.905) & -8.4 [-20.6, 3.1] (0.930) \\
Q10 \texttt{provider\_rec} & -3.7 [-13.2, 3.7] (0.839) & -3.8 [-12.9, 3.5] (0.856) & -6.5 [-19.5, 4.3] (0.876) & -7.7 [-20.7, 3.1] (0.916) & -8.8 [-22.4, 3.0] (0.927) & -9.8 [-22.8, 1.8] (0.951) \\
\textbf{Relationship} & -8.3 [-18.9, 0.6] (0.963) & \textbf{-8.3 [-18.4, -0.0]} (0.976) & -5.5 [-16.6, 4.6] (0.848) & -6.8 [-17.8, 3.4] (0.902) & -10.5 [-22.0, 0.7] (0.967) & \textbf{-11.4 [-22.2, -0.8]} (0.982) \\
Q11 \texttt{communication} & -11.9 [-30.5, 3.2] (0.933) & -12.0 [-30.6, 2.5] (0.945) & -5.7 [-19.4, 6.9] (0.809) & -7.3 [-21.4, 5.9] (0.866) & -12.6 [-27.9, 1.9] (0.953) & -13.5 [-28.4, 0.3] (0.972) \\
Q12 \texttt{dynamic} & -5.9 [-17.8, 1.3] (0.933) & -5.9 [-17.7, 1.1] (0.941) & -5.0 [-17.0, 5.5] (0.818) & -6.0 [-18.0, 5.1] (0.864) & -9.2 [-22.7, 2.5] (0.934) & -10.0 [-23.1, 1.8] (0.952) \\
Q13 \texttt{partner\_perspec} & -10.4 [-27.9, 1.9] (0.946) & -10.5 [-27.3, 1.2] (0.959) & -7.6 [-23.2, 6.4] (0.847) & -9.2 [-24.8, 5.0] (0.900) & -13.1 [-28.5, 1.7] (0.959) & \textbf{-14.2 [-28.8, -0.1]} (0.976) \\
Q14 \texttt{media\_rec} & -8.3 [-23.3, 4.8] (0.896) & -8.3 [-23.0, 4.3] (0.901) & -4.3 [-19.5, 12.7] (0.724) & -5.9 [-20.7, 11.3] (0.779) & -8.9 [-23.5, 7.1] (0.876) & -9.9 [-23.9, 5.4] (0.903) \\
Q15 \texttt{new\_sexual} & -4.8 [-15.9, 1.9] (0.911) & -4.9 [-15.3, 1.4] (0.927) & -4.8 [-16.1, 2.4] (0.885) & -5.5 [-17.3, 1.8] (0.917) & -8.6 [-21.7, 1.2] (0.955) & -9.2 [-21.7, 0.1] (0.973) \\
\textbf{All 15 scenarios} & -6.7 [-15.4, 1.6] (0.938) & -6.7 [-15.2, 1.0] (0.957) & -6.5 [-17.4, 3.8] (0.887) & -7.8 [-18.6, 2.4] (0.932) & -9.6 [-20.0, 0.9] (0.964) & \textbf{-10.5 [-20.3, -0.7]} (0.982) \\
\bottomrule
\end{tabular}%

\caption{Differences in chatbot use and willingness between LGBTQ+ and non-LGBTQ+ respondents, by domain and scenario, percentage points (LGBTQ+ minus non-LGBTQ+, new respondent). Each cell shows the mean difference, its 95\% credible interval in brackets and the probability that the difference is below zero in parentheses. Bold: the interval excludes zero. ``Used'' is actual use, ``Would'' is willingness among non-users, and ``UW'' combines the two. Domain rows average their 5 scenarios and the last row all 15. Frequency-adjusted values compare respondents with similar overall chatbot-use frequency; frequency was reported at the same time as scenario use, so this is not a baseline.}
\end{table*}
\end{landscape}

\begin{landscape}
\begin{table*}[p]
\centering
\footnotesize
\label{tab:B2c}
\begin{tabular}{lllllll}
\toprule
Domain / scenario & Used: unadjusted & Used: freq.-adj. & Would: unadjusted & Would: freq.-adj. & UW: unadjusted & UW: freq.-adj. \\
\midrule
\textbf{Mental health} & -6.7 [-17.5, 4.3] (0.887) & -4.2 [-14.4, 6.2] (0.784) & -10.8 [-22.5, 1.4] (0.958) & -9.8 [-21.4, 2.1] (0.949) & -11.8 [-24.4, 0.6] (0.968) & -9.3 [-20.8, 1.9] (0.949) \\
Q1 \texttt{mh\_issues} & -9.8 [-28.2, 7.3] (0.871) & -6.5 [-24.3, 9.5] (0.779) & -11.1 [-27.1, 4.8] (0.914) & -10.0 [-25.6, 5.8] (0.904) & -11.4 [-26.8, 2.5] (0.946) & -8.5 [-22.2, 3.8] (0.906) \\
Q2 \texttt{life\_change} & -4.0 [-18.9, 14.0] (0.712) & -1.4 [-16.1, 16.4] (0.597) & -12.9 [-28.2, 2.2] (0.955) & -11.6 [-27.0, 2.9] (0.942) & -10.9 [-26.0, 3.9] (0.928) & -8.4 [-22.5, 5.6] (0.884) \\
Q3 \texttt{therapist} & -4.2 [-15.8, 8.4] (0.787) & -2.5 [-13.5, 9.9] (0.694) & -10.4 [-23.1, 2.5] (0.944) & -9.6 [-22.6, 3.9] (0.927) & -11.5 [-25.2, 2.7] (0.945) & -9.4 [-22.7, 4.4] (0.915) \\
Q4 \texttt{conflict} & -2.0 [-15.3, 15.8] (0.645) & 0.3 [-12.6, 17.1] (0.524) & -10.5 [-23.6, 3.6] (0.936) & -9.4 [-22.3, 4.3] (0.914) & -9.1 [-23.6, 6.7] (0.882) & -6.7 [-20.5, 7.7] (0.827) \\
Q5 \texttt{body\_image} & -13.7 [-33.0, 2.6] (0.944) & -10.9 [-30.3, 5.3] (0.889) & -9.0 [-21.4, 2.8] (0.935) & -8.6 [-21.4, 3.2] (0.925) & \textbf{-16.2 [-33.2, -0.6]} (0.980) & -13.5 [-30.1, 1.5] (0.962) \\
\textbf{Sexual health} & -1.6 [-7.7, 5.8] (0.700) & -0.3 [-6.4, 7.3] (0.564) & -5.9 [-12.9, 1.7] (0.941) & -5.4 [-12.4, 2.1] (0.924) & -6.4 [-15.3, 3.4] (0.908) & -4.9 [-13.5, 4.4] (0.862) \\
Q6 \texttt{identity} & -5.1 [-16.3, 5.8] (0.843) & -3.4 [-14.5, 7.5] (0.754) & -5.4 [-13.2, 2.1] (0.929) & -5.1 [-12.8, 2.6] (0.914) & -9.0 [-20.7, 2.9] (0.937) & -7.1 [-18.7, 4.4] (0.898) \\
Q7 \texttt{sti} & -3.4 [-14.1, 8.7] (0.765) & -1.8 [-12.1, 10.6] (0.671) & -8.4 [-19.8, 4.1] (0.918) & -7.6 [-18.9, 5.0] (0.899) & -9.5 [-22.4, 4.9] (0.915) & -7.6 [-20.2, 6.0] (0.882) \\
Q8 \texttt{gender\_care} & 2.5 [-5.9, 18.1] (0.413) & 3.6 [-4.8, 20.2] (0.329) & -5.0 [-12.1, 2.1] (0.933) & -4.7 [-11.9, 2.4] (0.920) & -2.3 [-12.9, 12.8] (0.688) & -1.0 [-11.6, 14.4] (0.620) \\
Q9 \texttt{side\_effects} & 0.9 [-7.2, 14.3] (0.521) & 1.9 [-5.9, 15.8] (0.424) & -5.0 [-12.4, 2.3] (0.923) & -4.6 [-12.0, 2.9] (0.906) & -3.8 [-14.0, 9.7] (0.773) & -2.4 [-12.6, 11.1] (0.707) \\
Q10 \texttt{provider\_rec} & -2.7 [-10.0, 5.1] (0.810) & -1.7 [-9.2, 6.0] (0.718) & -5.7 [-14.7, 4.9] (0.890) & -5.1 [-14.0, 5.5] (0.869) & -7.4 [-17.9, 4.5] (0.904) & -6.1 [-16.2, 5.3] (0.876) \\
\textbf{Relationship} & -4.4 [-11.9, 3.4] (0.876) & -2.8 [-10.1, 4.6] (0.781) & -7.4 [-15.5, 1.3] (0.954) & -6.8 [-14.9, 1.6] (0.944) & -9.5 [-19.3, 0.7] (0.966) & -7.7 [-17.0, 1.8] (0.946) \\
Q11 \texttt{communication} & -7.8 [-23.6, 7.6] (0.853) & -5.2 [-21.4, 10.5] (0.752) & -7.2 [-17.8, 3.6] (0.916) & -6.8 [-17.1, 4.3] (0.904) & -11.2 [-26.3, 3.5] (0.931) & -8.6 [-23.5, 6.0] (0.882) \\
Q12 \texttt{dynamic} & -3.6 [-11.6, 3.0] (0.874) & -2.7 [-10.5, 3.8] (0.808) & -6.1 [-14.8, 2.6] (0.923) & -5.7 [-14.3, 3.2] (0.909) & -8.6 [-19.0, 1.8] (0.950) & -7.4 [-17.5, 2.8] (0.934) \\
Q13 \texttt{partner\_perspec} & -2.6 [-13.8, 11.5] (0.706) & -0.8 [-11.9, 13.1] (0.595) & -9.4 [-21.2, 2.9] (0.940) & -8.5 [-20.2, 3.3] (0.926) & -9.6 [-23.1, 4.7] (0.909) & -7.5 [-20.4, 6.3] (0.868) \\
Q14 \texttt{media\_rec} & -6.1 [-19.1, 6.9] (0.841) & -4.2 [-17.3, 8.4] (0.756) & -9.7 [-21.6, 2.7] (0.942) & -8.9 [-21.4, 3.5] (0.924) & -12.0 [-25.8, 2.1] (0.955) & -9.9 [-23.4, 3.5] (0.927) \\
Q15 \texttt{new\_sexual} & -2.1 [-8.6, 5.0] (0.792) & -1.3 [-7.7, 5.8] (0.699) & -4.5 [-11.2, 0.9] (0.950) & -4.3 [-11.2, 1.2] (0.942) & -6.1 [-14.6, 2.3] (0.931) & -5.1 [-13.5, 3.2] (0.905) \\
 \textbf{All 15 scenarios} & -4.2 [-11.5, 3.5] (0.867) & -2.4 [-9.4, 4.9] (0.750) & -8.0 [-16.5, 1.0] (0.959) & -7.4 [-15.8, 1.5] (0.948) & -9.2 [-19.0, 0.9] (0.961) & -7.3 [-16.4, 2.0] (0.941) \\
\bottomrule
\end{tabular}
\caption{Differences in chatbot use and willingness between trans and non-trans respondents, by domain and scenario, percentage points (trans minus non-trans, new respondent). Each cell shows the mean difference, its 95\% credible interval in brackets and the probability that the difference is below zero in parentheses. Bold: the interval excludes zero. ``Used'' is actual use, ``Would'' is willingness among non-users, and ``UW'' combines the two. Domain rows average their 5 scenarios and the last row all 15. Frequency-adjusted values compare respondents with similar overall chatbot-use frequency; frequency was reported at the same time as scenario use, so this is not a baseline.}
\end{table*}
\end{landscape}

\begin{landscape}
\begin{table*}[p]
\centering
\footnotesize
\label{tab:B2d}
\begin{tabular}{lllllll}
\toprule
Domain / scenario & Used: unadjusted & Used: freq.-adj. & Would: unadjusted & Would: freq.-adj. & UW: unadjusted & UW: freq.-adj. \\
\midrule
\textbf{Mental health} & -4.1 [-15.8, 8.0] (0.752) & -1.7 [-13.0, 9.8] (0.612) & -7.4 [-20.7, 6.4] (0.859) & -6.2 [-19.2, 7.1] (0.820) & -7.9 [-22.0, 5.7] (0.872) & -5.4 [-18.5, 6.6] (0.793) \\
Q1 \texttt{mh\_issues} & -7.5 [-27.6, 10.6] (0.782) & -4.3 [-23.5, 12.6] (0.673) & -7.8 [-25.5, 10.1] (0.811) & -6.5 [-23.3, 10.4] (0.776) & -8.5 [-25.2, 6.4] (0.858) & -5.6 [-20.8, 7.9] (0.778) \\
Q2 \texttt{life\_change} & -1.1 [-17.0, 17.3] (0.570) & 1.5 [-14.4, 19.8] (0.444) & -9.5 [-26.7, 7.2] (0.871) & -7.8 [-24.5, 8.0] (0.830) & -7.1 [-23.9, 8.7] (0.809) & -4.5 [-19.9, 10.2] (0.720) \\
Q3 \texttt{therapist} & -2.3 [-14.1, 10.6] (0.674) & -0.6 [-12.5, 12.5] (0.558) & -6.7 [-20.6, 7.7] (0.829) & -5.6 [-19.4, 8.6] (0.790) & -7.2 [-22.1, 8.3] (0.826) & -5.1 [-19.5, 9.5] (0.754) \\
Q4 \texttt{conflict} & 0.7 [-13.4, 19.1] (0.500) & 2.8 [-10.7, 20.0] (0.385) & -7.0 [-21.2, 8.4] (0.825) & -5.7 [-19.8, 9.3] (0.784) & -4.8 [-20.6, 12.0] (0.729) & -2.5 [-17.7, 12.6] (0.630) \\
Q5 \texttt{body\_image} & -10.4 [-30.2, 6.7] (0.872) & -7.8 [-27.9, 9.4] (0.788) & -6.0 [-19.4, 7.2] (0.828) & -5.3 [-18.2, 7.4] (0.797) & -12.0 [-29.7, 4.6] (0.915) & -9.3 [-26.6, 6.3] (0.867) \\
 \textbf{Sexual health} & -0.1 [-6.9, 7.8] (0.541) & 1.1 [-5.3, 8.9] (0.396) & -3.6 [-11.4, 5.2] (0.816) & -3.0 [-10.7, 5.2] (0.773) & -3.2 [-13.0, 7.4] (0.735) & -1.6 [-10.9, 8.3] (0.634) \\
Q6 \texttt{identity} & -3.7 [-16.0, 8.4] (0.744) & -2.0 [-14.0, 9.7] (0.643) & -3.4 [-11.9, 5.3] (0.813) & -2.9 [-11.2, 5.5] (0.777) & -6.1 [-19.1, 7.1] (0.833) & -4.2 [-16.7, 8.1] (0.750) \\
Q7 \texttt{sti} & -1.4 [-12.0, 11.0] (0.621) & 0.2 [-10.4, 12.8] (0.508) & -5.3 [-17.9, 8.6] (0.801) & -4.3 [-17.0, 9.3] (0.758) & -5.5 [-19.6, 9.6] (0.777) & -3.5 [-17.2, 10.9] (0.697) \\
Q8 \texttt{gender\_care} & 3.8 [-5.0, 19.4] (0.307) & 4.8 [-3.8, 21.5] (0.224) & -3.1 [-10.5, 4.7] (0.821) & -2.5 [-9.6, 5.1] (0.785) & 0.6 [-10.7, 16.3] (0.506) & 2.0 [-8.8, 17.5] (0.424) \\
Q9 \texttt{side\_effects} & 2.1 [-6.3, 15.6] (0.401) & 3.2 [-5.1, 17.5] (0.306) & -3.0 [-10.8, 5.5] (0.804) & -2.5 [-10.1, 5.7] (0.768) & -0.8 [-11.6, 13.1] (0.592) & 0.5 [-9.9, 14.1] (0.503) \\
Q10 \texttt{provider\_rec} & -1.5 [-9.0, 6.5] (0.692) & -0.6 [-7.8, 7.6] (0.592) & -3.1 [-12.7, 8.3] (0.751) & -2.5 [-12.0, 8.6] (0.720) & -4.1 [-15.2, 8.4] (0.769) & -2.8 [-13.4, 9.2] (0.702) \\
\textbf{Relationship} & -2.5 [-10.5, 5.8] (0.732) & -0.9 [-8.5, 7.2] (0.594) & -4.8 [-14.1, 5.1] (0.844) & -4.1 [-13.1, 5.4] (0.810) & -5.9 [-16.9, 5.4] (0.854) & -4.0 [-14.3, 6.2] (0.782) \\
Q11 \texttt{communication} & -4.8 [-21.4, 11.7] (0.723) & -2.2 [-18.4, 13.7] (0.607) & -4.6 [-15.8, 7.3] (0.799) & -3.8 [-15.0, 8.2] (0.767) & -7.0 [-22.9, 8.7] (0.814) & -4.4 [-19.5, 10.9] (0.712) \\
Q12 \texttt{dynamic} & -2.3 [-9.9, 4.2] (0.778) & -1.5 [-8.8, 5.2] (0.688) & -3.8 [-13.2, 6.0] (0.802) & -3.2 [-12.4, 6.7] (0.770) & -5.5 [-16.4, 5.7] (0.846) & -4.2 [-14.7, 6.6] (0.791) \\
Q13 \texttt{partner\_perspec} & -0.3 [-11.7, 14.1] (0.557) & 1.4 [-9.6, 15.6] (0.440) & -6.0 [-18.6, 7.4] (0.830) & -5.1 [-17.5, 8.0] (0.788) & -5.2 [-19.5, 10.2] (0.758) & -3.2 [-17.1, 11.1] (0.678) \\
Q14 \texttt{media\_rec} & -3.9 [-17.5, 9.8] (0.731) & -1.9 [-15.4, 11.6] (0.622) & -6.6 [-20.2, 7.3] (0.838) & -5.6 [-19.0, 7.8] (0.800) & -8.1 [-23.3, 7.1] (0.853) & -5.8 [-20.6, 8.8] (0.780) \\
Q15 \texttt{new\_sexual} & -1.0 [-7.4, 5.9] (0.663) & -0.2 [-6.4, 7.1] (0.560) & -3.0 [-10.0, 3.3] (0.849) & -2.5 [-9.4, 3.5] (0.812) & -3.7 [-12.5, 5.2] (0.812) & -2.6 [-11.1, 6.1] (0.746) \\
\textbf{All 15 scenarios} & -2.2 [-10.3, 6.2] (0.711) & -0.5 [-8.1, 7.6] (0.555) & -5.2 [-15.0, 5.2] (0.852) & -4.4 [-13.9, 5.5] (0.813) & -5.7 [-16.8, 5.6] (0.839) & -3.7 [-13.9, 6.5] (0.759) \\
\bottomrule
\end{tabular}
\caption{Differences in chatbot use and willingness between trans and non-trans respondents among LGBTQ+ respondents, by domain and scenario, percentage points (trans minus non-trans, both LGBTQ+, new respondent). Each cell shows the mean difference, its 95\% credible interval in brackets and the probability that the difference is below zero in parentheses. Bold: the interval excludes zero. ``Used'' is actual use, ``Would'' is willingness among non-users, and ``UW'' combines the two. Domain rows average their 5 scenarios and the last row all 15. Frequency-adjusted values compare respondents with similar overall chatbot-use frequency; frequency was reported at the same time as scenario use, so this is not a baseline. The trans vs non-trans LGBTQ+ comparison comes from a model fitted to all 73 respondents with separate LGBTQ+, trans and US-residence terms; for each of the 60 LGBTQ+ respondents it predicts the probability twice, once as trans and once as non-trans, keeping their actual US residence and (in the frequency-adjusted columns) their overall chatbot-use frequency, and averages the difference over the 60. The model treats the trans difference as the same for LGBTQ+ and non-LGBTQ+ respondents; this has little influence here because only 1 of the 21 trans respondents is not LGBTQ+.}
\end{table*}
\end{landscape}

\begin{landscape}
\begin{table}[htbp]
\centering
\footnotesize
\label{tab:b1-prior-sensitivity}
\begin{tabular}{lcccc}
\toprule
Specification & LGBTQ+ overall & LGBTQ+ relationship & Trans overall & Trans relationship \\
\midrule
Main analysis & -8.5 [-18.6, 0.5] & \textbf{-10.0 [-21.7, -0.4]} & -5.0 [-12.4, 2.5] & -5.2 [-12.7, 2.4] \\
Fixed-effect prior: wide identity slope N(0,1.5) + item-slope SD exp(1) & \textbf{-10.0 [-20.4, -0.2]} & \textbf{-12.0 [-24.3, -1.4]} & -5.8 [-13.3, 2.2] & -6.2 [-14.2, 2.1] \\
Structure: separate fixed identity slope per domain & -6.5 [-14.3, 0.9] & \textbf{-15.3 [-26.8, -4.5]} & -4.0 [-10.1, 2.7] & -6.4 [-14.0, 1.8] \\
Hierarchical prior: Item-slope SD exp(8) (4x tighter) & -8.5 [-18.3, 0.3] & -8.5 [-18.5, 0.0] & -5.1 [-12.1, 2.7] & -4.8 [-11.8, 2.5] \\
Hierarchical prior: Item-slope SD exp(0.5) (4x looser) & -8.4 [-18.4, 0.3] & \textbf{-10.9 [-22.9, -0.8]} & -5.3 [-12.4, 2.5] & -5.7 [-13.1, 2.3] \\
Hierarchical prior: Item-intercept SD exp(8) (4x tighter) & -8.5 [-18.1, 0.3] & \textbf{-10.0 [-21.3, -0.5]} & -5.1 [-12.3, 2.5] & -5.3 [-12.6, 2.5] \\
Hierarchical prior: Item-intercept SD exp(0.5) (4x looser) & -8.4 [-18.2, 0.6] & \textbf{-10.0 [-21.5, -0.5]} & -5.1 [-12.2, 2.6] & -5.3 [-12.5, 2.5] \\
Hierarchical prior: Person SD exp(4) (4x tighter) & -8.5 [-17.9, 0.1] & \textbf{-9.9 [-21.3, -0.7]} & -5.1 [-12.0, 2.0] & -5.3 [-12.5, 2.0] \\
Hierarchical prior: Person SD exp(0.25) (4x looser) & -8.4 [-18.1, 0.3] & \textbf{-9.9 [-21.5, -0.5]} & -5.1 [-12.4, 2.8] & -5.2 [-12.8, 2.7] \\
Hierarchical prior: All SDs half-Student-t(3,0,1) & -8.6 [-17.8, 0.2] & \textbf{-10.9 [-22.7, -1.2]} & -5.1 [-12.2, 2.4] & -5.5 [-13.1, 2.5] \\
Country: + US-residence indicator (standardised to sample mix) & -6.9 [-17.0, 2.5] & -8.5 [-20.4, 1.5] & -3.1 [-11.8, 6.3] & -3.5 [-12.0, 5.6] \\
Country: + country-of-residence random intercept (new person, new country) & -7.8 [-17.7, 1.9] & -9.5 [-21.3, 0.8] & -4.3 [-11.7, 4.5] & -4.6 [-12.1, 4.2] \\
Country: single estimable residence stratum (LGBTQ+: outside US; trans: US) & -7.0 [-17.1, 2.5] & -9.0 [-21.3, 1.2] & -6.5 [-17.0, 3.8] & -5.8 [-16.4, 3.6] \\
\bottomrule
\end{tabular}%
\caption{Response differences across prior specifications: overall and for relationship scenarios. Values are posterior means with 95\% credible intervals, in percentage points. Bold intervals exclude zero.}
\end{table}
\end{landscape}

\begin{landscape}
\begin{table}[htbp]
\centering
\footnotesize
\begin{tabular}{llcc}
\toprule
Target population & Outcome & Original prior: identity/residence SD 1, frequency SD 0.25 & Stronger prior: identity/residence SD 0.5, frequency SD 0.125 \\
\midrule
All LGBTQ+ respondents & Actual use & -0.48 [-8.12, 7.59] & -0.21 [-7.61, 7.63] \\
All LGBTQ+ respondents & Willingness among non-users & -4.40 [-13.85, 5.51] & -4.14 [-13.82, 5.37] \\
All LGBTQ+ respondents & Used or would use & -3.67 [-13.89, 6.54] & -3.29 [-13.41, 6.15] \\
US LGBTQ+ respondents & Actual use & -0.48 [-7.86, 6.76] & -0.31 [-7.70, 6.56] \\
US LGBTQ+ respondents & Willingness among non-users & -3.98 [-13.37, 5.01] & -3.85 [-13.34, 4.84] \\
US LGBTQ+ respondents & Used or would use & -3.44 [-13.32, 6.21] & -3.23 [-12.93, 5.68] \\
\bottomrule
\end{tabular}%
\caption{Prior sensitivity of the joint, overall-frequency-adjusted use/willingness model: trans minus non-trans response differences among LGBTQ+ respondents, averaged over 15 scenarios and standardised to the target population's residence and frequency profiles. Values are posterior means [95\% CrI] in percentage points. The stated SDs refer to class-membership coefficient priors. Stronger-prior estimates are obtained by importance reweighting the original fit.}
\end{table}
\end{landscape}

\clearpage

\end{document}